\documentclass[twocol]{ametsoc}

\journal{jcli}
\bibpunct{(}{)}{;}{a}{}{,}

\usepackage{text_shortcuts,math_shortcuts,phys_shortcuts}

\title{Symmetric and antisymmetric components of polar-amplified warming}

\authors{
  Spencer A. Hill
  \correspondingauthor{Spencer Hill, 207B Oceanography, Lamont-Doherty Earth Observatory, 61 Route 9W, Palisades, NY 10964}
  \thanks{Current Affiliation: Program in Atmospheric and Oceanic Sciences, Princeton University, Princeton, New Jersey}
}

\affiliation{Lamont-Doherty Earth Observatory, Columbia University, Palisades, New York}
\email{spencerh@princeton.edu}

\extraauthor{Natalie J. Burls}
\extraaffil{Center for Ocean-Land-Atmosphere Studies, Department of Atmospheric, Oceanic, and Earth Sciences, George Mason University, Fairfax, Virginia}

\extraauthor{Alexey Fedorov}
\extraaffil{Department of Earth \& Planetary Sciences, Yale University, New Haven, Connecticut, and LOCEAN/IPSL, Sorbonne University, Paris, France}

\extraauthor{Timothy M. Merlis}
\extraaffil{Department of Atmospheric and Oceanic Sciences, McGill University, Montreal, Quebec, Canada}

\abstract{
\coo/-forced surface warming in general circulation models (GCMs) is initially polar-amplified in the Arctic but not Antarctic---a largely hemispherically antisymmetric signal.  Nevertheless, we show in CESM1 and eleven LongRunMIP GCMs that the hemispherically symmetric component of global-mean-normalized, zonal-mean warming (\(\Tsn\)) under \ftcoo/ changes weakly or becomes moderately more polar-amplified from the first decade to near-equilibrium.  Conversely, the antisymmetric warming component (\(\Tan\)) weakens with time in all models, moderately in some including FAMOUS but effectively vanishing in others including CESM1.  We explore mechanisms underlying the robust \(\Tsn\) behavior with a diffusive moist energy balance model (MEBM), which given radiative feedback parameter (\(\lambda\)) and ocean heat uptake (\(\ohu\)) fields diagnosed from CESM1 adequately reproduces the CESM1 \(\Tsn\) and \(\Tan\) fields.  In further MEBM simulations perturbing \(\lambda\) and \(\ohu\), \(\Tsn\) is sensitive to their symmetric components only, and more to that of \(\lambda\).  A three-box, two-timescale model fitted to FAMOUS and CESM1 reveals a curiously short Antarctic fast-response timescale in FAMOUS.  In additional CESM1 simulations spanning a broader range of forcings, \(\Tsn\) changes modestly across 2-\stcoo/, and \(\Tsn\) in a Pliocene-like simulation is more polar-amplified but likewise approximately time-invariant.  Determining the real-world relevance of these behaviors---which imply that a surprising amount of information about near-equilibrium polar amplification emerges within decades---merits further study.
}

\begin{document}
\maketitle

\section{Introduction}
\label{sec:intro}

\begin{figure}[h]
\centering
\includegraphics[width=0.5\textwidth]{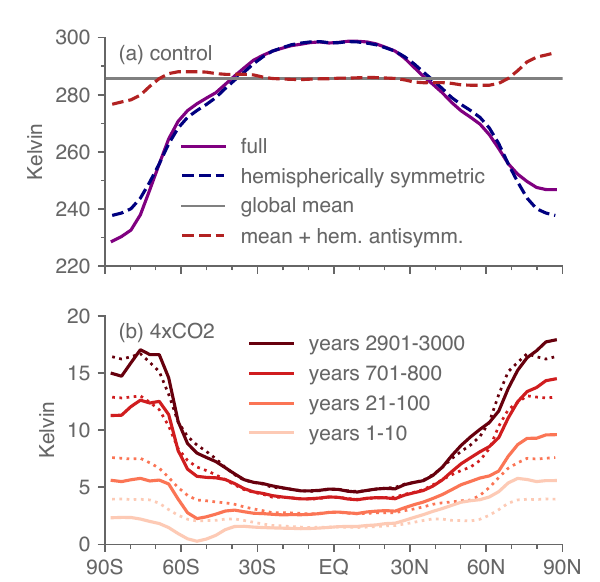}
\caption{(a) Climatological annual-mean, zonal-mean surface air temperature in a preindustrial control simulation, the hemispherically symmetric component thereof, the global mean thereof, and the global mean plus the hemispherically antisymmetric component, as indicated in the legend.  The simulation was performed in a low-resolution configuration of CESM version 1.0.4, with results averaged over years 701-800.  (b) Anomalous zonal-mean surface air temperature in an abrupt \ftcoo/ simulation in the same model, averaged over four time periods as indicated in the legend.  Solid curves are the full fields, and dotted curves are the corresponding symmetric component.  Both panels have units Kelvin.}
\label{fig:tsurf-cont-4x}
\end{figure}

Climatological zonal-mean surface temperatures decrease from the equator toward both poles, a hemispherically symmetric signature much larger than the antisymmetric deviations therefrom.  By symmetric or antisymmetric we refer to the average or difference, respectively, of each latitude with its mirror about the equator: for a given field \(\chi\), \(\chi(\lat)=\chi\sym(\lat)+\chi\asym(\lat)\), where \(\lat\) is latitude, \(\chi\sym\equiv\frac{1}{2}[\chi(\lat)+\chi(-\lat)]\) is the symmetric component, and \(\chi\asym\equiv\frac{1}{2}[\chi(\lat)-\chi(-\lat)]\) is the antisymmetric component.
Fig.~\ref{fig:tsurf-cont-4x}(a) illustrates this via a preindustrial control simulation in the Community Earth System Model version 1.0.4 (henceforth CESM1) general circulation model (GCM) whose formulation will be described below.  Evidently, the symmetric annual-mean forcing of insolation and approximately uniform forcing of \coo/ and other well-mixed greenhouse gases outweigh the antisymmetric components of Earth's ocean basins, orography, sea ice, clouds, and atmospheric and oceanic circulations.

Conversely, \coo/-forced zonal-mean surface warming---henceforth simply \(T\)---starts out appreciably antisymmetric: prevailing Southern Ocean upwelling \citep[\eg/][]{armour_time-varying_2013,marshall_oceans_2015} impedes Antarctic warming for decades \citep[likely reinforced by resulting changes in local lapse rates and clouds;][]{senior_time-dependence_2000,rugenstein_equilibrium_2020}, while weakly negative to slightly positive radiative feedbacks in northern high latitudes \citep[\eg/][]{stuecker_polar_2018} among other processes \citep{feldl_atmospheric_2017,russotto_polar_2020,henry_decomposing_2021} promote Arctic warming.  This fast response typically gives way to a more symmetric warming pattern over subsequent centuries \citep{held_probing_2010}, with Antarctic warming partially catching up to the Arctic in century-scale CMIP5 \citep{andrews_dependence_2015} and CMIP6 \citep{dong_intermodel_2020} simulations.  On longer timescales, polar amplification is comparable in the two hemispheres in multi-millennial simulations in fully coupled GCMs \citep[\eg/][]{danabasoglu_equilibrium_2009,li_deep-ocean_2013,rugenstein_longrunmip_2019} and in runs to equilibrium in slab-ocean GCMs \citep[\eg/][]{manabe_transient_1991,armour_time-varying_2013} and diffusive moist energy balance models (MEBMs) \citep[\eg/][]{merlis_simple_2018,armour_meridional_2019}.  Fig.~\ref{fig:tsurf-cont-4x}(b) illustrates these behaviors via \(T\) and \(\Tsym\) from an abrupt \ftcoo/ simulation in CESM1 over each of four time periods (years 1-10, 21-100, 701-800, and 2901-3000): \(T\) and \(\Tsym\) differ markedly in the first decade when \(T\asym\) is largest but gradually become more similar, with \(T\approx \Tsym\) and \(T\asym\approx 0\) to first approximation in the final period.

Nevertheless, past studies indicate that \(T\) normalized by its global average---henceforth \(T^*\)---partially collapses toward a shared pattern at different timescales \citep[\cf/ Fig.~4a of][]{armour_time-varying_2013}.
Such pattern scaling \citep[\eg/][]{tebaldi_pattern_2014} also holds for zonally varying surface temperature responses across \coo/ values \citep{heede_time_2020}.
Essentially, the present study combines pattern scaling and the symmetric/antisymmetric decomposition for \(T\) under \ftcoo/, arguing that GCM-simulated \(\Tsn\) changes surprisingly little from the first decade to near-equilibrium while \(\Tan\) weakens markedly.  Taken at face value, this would imply that polar amplification (defined as the ratio of polar-cap to globally averaged warming) at near-equilibrium in a given GCM can be meaningfully constrained from a single decade of forced change.

We found these behaviors somewhat inadvertently in the aforementioned \ftcoo/ simulation in CESM1 (which is described along with the other models and methodological choices in Section~\ref{sec:methods}), and this manuscript constitutes an attempt to better understand them.  To assess their robustness, we analyze eleven additional GCMs from the LongRunMIP \citep{rugenstein_longrunmip_2019} repository (Section~\ref{sec:longrunmip}).  To clarify their underlying physical mechanisms, we use an MEBM to first emulate the results from CESM1 and then identify the predominant factors determining \(\Tsn\) (Section~\ref{sec:mebm}).  To better understand their differences across these GCMs, we fit a three-box, two-timescale model to two end-members of these twelve, CESM1 and FAMOUS (Section~\ref{sec:box-model}).  And to assess their relevance across different forcings, we analyze 2, 8, and \stcoo/ simulations and the aforementioned Pliocene-like simulation in CESM1 (Section~\ref{sec:co2}).  We conclude with summary and discussion (Section~\ref{sec:conc}), including comparison to the more traditional approach of studying polar amplification in either cap separately, potential means of further testing these behaviors, and implications for the real climate system.



\section{Methods}
\label{sec:methods}

\subsection{LongRunMIP and CESM1 \ftcoo/ simulations}
LongRunMIP \citep{rugenstein_longrunmip_2019} comprises increased-\coo/ simulations from CMIP5-class GCMs spanning from one thousand to several thousand years.  We analyze eleven of the twelve available with 1,000+ year integrations under \ftcoo/, which are listed in Table~\ref{table:longrunmip}.  CESM104 from LongRunMIP was omitted because it is nearly the same as the above-noted CESM1 that we analyze separately.  Nine of the LongRunMIP models ran under an abrupt \ftcoo/ and two under a 1\% increase per year to \ftcoo/.  The latter two (ECHAM5MPIOM and MIROC32) also ran shorter abrupt \ftcoo/ simulations, and so for the first century when forcing and global-mean warming are modest under 1\%/yr we use the abrupt \ftcoo/ simulation, switching to the 1\% to \ftcoo/ simulation for subsequent periods.  Output was available regridded to a common 2.5\(\times\)2.5\degr{} grid \citep[\cf/ Table~2 of][]{rugenstein_longrunmip_2019}.

\begin{table}
  \begin{center}
    \begin{tabular}{lrll}
      \hline\hline
      Model&Control dur.&701-800 sim.&2901-3000 sim.\\
      \hline
      CCSM3 & 1530 & \ftcoo/ & none\\
      CNRMCM61 & 2000 & \ftcoo/ & none\\
      ECHAM5MPIOM & 100 & 1\%\ftcoo/ & 1\%\ftcoo/\\
      FAMOUS & 3000 & \ftcoo/ & \ftcoo/\\
      GISSE2R & 5225 & \ftcoo/ & \ftcoo/\\
      HadCM3L & 1000 & \ftcoo/ & none\\
      HadGEM2 & 239 & \ftcoo/ & none\\
      IPSLCM5A & 1000 & \ftcoo/ & none\\
      MIROC32 & 680 & 1\%\ftcoo/ & none\\
      MPIESM11 & 2000 & \ftcoo/ & \ftcoo/\\
      MPIESM12 & 1237 & \ftcoo/ & none\\
      \hline
    \end{tabular}
  \end{center}
  \caption{Details of the LongRunMIP models and simulations used.  Columns, from left to right: Model name following \citet{rugenstein_longrunmip_2019} conventions; control simulation duration in years; simulation for which data over years 701-800 is taken; and simulation for which data over years 2901-3000 is taken (or ``none'' if not available for any simulation in that model)}
  \label{table:longrunmip}
\end{table}

We include with the LongRunMIP models the 3,000-yr \ftcoo/ simulation in CESM1 referred to in the Introduction.  This is version 1.0.4 of the model in its low-resolution configuration \citep{shields_low-resolution_2012}.  It consists of the Community Atmosphere Model, version 4 with its spectral dynamical core truncated at T31 resolution (\(\sim\)\(3.75^\circ\times3.75^\circ\)) and with 26 vertical levels coupled to the Parallel Ocean Program version 2 (POP2) with \(\sim\)3\degr{} horizontal resolution and 60 vertical levels.

We focus on temporal averages over four time periods \citep[similar to those of][]{armour_time-varying_2013}: years 1-10 and 21-100 (during which both the atmosphere and ocean are rapidly responding), 701-800 (during which the atmosphere is in a nearly statistically steady state but the ocean remains slowly varying), and 2901-3000 (at which time the deep ocean has nearly equilibrated).  All twelve GCMs extend through year 800 and five through year 3000.  We refer to the final period as near-equilibrium, recognizing that the climate response would likely meaningfully evolve beyond three millennia in most models given the deep ocean's multi-millennial, diffusive equilibration timescale \citep{jansen_transient_2018}; the box model in Section~\ref{sec:box-model} highlights this.

We account for climate drift in each model's preindustrial control simulation as follows.  For the eleven LongRunMIP GCMs, for each time period we compute anomalies as the difference between the \ftcoo/ simulation and the control at that time period if the control simulation extends that long.  Otherwise, we subtract an average over the entire control simulation.  For CESM1, drift in the control simulation is modest relative to the forced temperature responses, and so for convenience we report anomalies in all periods as differences with the control averaged over years 701-800.  All major results presented are insensitive to reasonable methodological choices regarding control drift.

The five GCMs extending to year 3,000 include the three farthest on the ends of the full twelve-GCM distribution at years 701-800: FAMOUS on one end (highest global-mean warming, and second-weakest changes in both the symmetric and antisymmetric polar amplification indices defined below), versus GISSE2R and CESM1 (respectively, lowest and third lowest mean warming, largest and second-largest increase in symmetric amplification, and second-largest and largest decrease in antisymmetric amplification) on the other.  Most likely then this subset usefully approximates the range generated by all twelve models had the others also run to year 3,000.

\subsection{Moist energy balance model}
To clarify the processes determining the GCM \(\Tsn\) and \(\Tan\) behaviors of our interest, we use a highly idealized diffusive moist energy balance model (MEBM).  MEBMs have been a useful simplified modeling framework for emulating the warming pattern in comprehensive GCMs \citep{hwang_coupling_2011,bonan_sources_2018} and developing theory for the spatial pattern of warming \citep{flannery_energy_1984,rose_dependence_2014,roe_remote_2015,merlis_simple_2018,russotto_polar_2020}, and we pursue both purposes here.
In short, the MEBM is forced with a realistic, time-invariant estimate of \ftcoo/ radiative forcing and, for each of the four selected time periods, input fields taken from the CESM1 \ftcoo/ simulation---either unmodified or perturbed as will be described in Section~\ref{sec:mebm}.

The MEBM's governing equation is
\begin{equation}
\label{eq:mebm}
\mathcal{C} \partial_t T(\lat) = \mathcal{F}(\lat) + \lambda(\lat) T(\lat) - \ohu(\lat) + \mathcal{D} \nabla^2 h(\lat),
\end{equation}
where \(\mathcal{C}\) is the surface layer heat capacity, \(T\) is anomalous surface temperature, \(\mathcal{F}\) is the imposed radiative forcing, \(\lambda\) is the radiative feedback parameter, \(\ohu\) is the anomalous net surface flux (signed positive downward; also known as ocean heat uptake), \(\mathcal{D}\) is the spatially uniform diffusivity, and \(h\) is surface moist static energy (MSE).  In words, the time tendency of the heat content of the surface layer (LHS) is determined by the combined effect of (RHS terms, left to right) the imposed radiative forcing (which is identical across all MEBM runs), a radiative restoring term encompassing the net effect of all TOA radiative feedbacks and that varies linearly with the surface temperature anomaly, an imposed ocean heat uptake field, and the convergence of the anomalous column-integrated MSE flux, approximated as downgradient diffusion of surface MSE.  The MEBM numerics and the calculations of each RHS term are conventional and detailed in the Appendix.

\subsection{Additional CESM1 simulations under different forcings}
To assess how robust the behaviors of \(\Tsn\) and \(\Tan\) are to forcings other than \ftcoo/, we also analyze instantaneous 2, 8, and \stcoo/ simulations in CESM1, as well as the Pliocene-like simulation mentioned in the Introduction \citep{burls_simulating_2014}.  In the latter, atmospheric composition remains preindustrial, but---only in shortwave radiative transfer calculations---liquid water path is decreased by 240\% poleward of 15\degr{} in both hemispheres, while both ice and liquid water paths are increased by 60\% within 15\degr{}S-15\degr{}N.  This increases the albedo of the deep tropical band, promoting local cooling, but decreases the albedo elsewhere, promoting warming \citep{burls_what_2014,fedorov_tightly_2015}.  Each spans 3,000 years.

\subsection{Physical meaning of symmetric/antisymmetric decomposition}
Arguably, the decomposition of \(T\) into a sum and difference of its mirror values about the equator---though always permissible mathematically---gains physical meaning only to the extent that mirror latitudes influence one another.  Otherwise, if for example each latitude was in local radiative-convective equilibrium independent of all others, summing or differencing about the equator merely convolves two independent signals.  But it is well understood that perturbed atmospheric and oceanic energy flux divergences do strongly influence polar amplification \citep[\eg/][]{alexeev_polar_2012,armour_meridional_2019,henry_decomposing_2021}, mitigating this concern, at least over sufficiently long timescales.

Nevertheless, a corollary is that this decomposition becomes physically meaningful only beyond the timescale over which a given latitude plausibly influence its mirror.  For example, while \citet{previdi_arctic_2020} argue convincingly that Arctic amplification emerges in a matter of months after imposed \coo/ forcing, for our purposes this Arctic signal is unlikely communicated to the opposite pole on such a sub-annual timescale.  \citet{shin_how_2021} show that, in an aquaplanet GCM with radiative forcing confined to one hemisphere's extratropics, local warming is communicated to the opposite polar cap through a multi-step circulation adjustment, manifesting in surface warming over \(\sim\)5-10 years.  For non-aquaplanets, zonal asymmetries plausibly yield teleconnections mediated by Rossby waves that could potentially transmit the signal across the tropics more rapidly \citep{ding_tropical_2014}; nevertheless we take the \citet{shin_how_2021} result as \emph{a posteriori} justification for our choice of the first decade as the earliest and shortest period analyzed.

Though a few prior studies have applied the symmetric/antisymmetric decomposition to related properties of the atmospheric energy budget, to our knowledge none have applied it to surface warming itself.  \citet{frierson_extratropical_2012} use the antisymmetric component of zonal-mean net energetic forcing of the atmosphere to interpret tropical precipitation and atmospheric energy fluxes under doubled \coo/.  In terms of hemispheric averages, observations and GCMs exhibit considerable symmetry in top-of-atmosphere (TOA) albedo climatologically \citep{voigt_observed_2013,stephens_albedo_2015} and in GCMs under hemispherically antisymmetric external forcing \citep{voigt_compensation_2014}.

\subsection{Amplification indices}
As quantitative bulk measures of \(\Tsn\) and \(\Tan\),
we start with conventional indices of polar amplification: Arctic amplification is the ratio of \(T\) averaged over 60-90\degr{}N to its global-mean, and Antarctic amplification the same but using 60-90\degr{}S.  We then define symmetric (\(\pasym\)) and antisymmetric (\(\paasym\)) polar amplification indices as the average or half the difference of the Arctic and Antarctic indices, respectively.

\section{\ftcoo/ results in GCMs}
\label{sec:longrunmip}

\begin{figure*}[h]
\centering
\includegraphics[width=\textwidth]{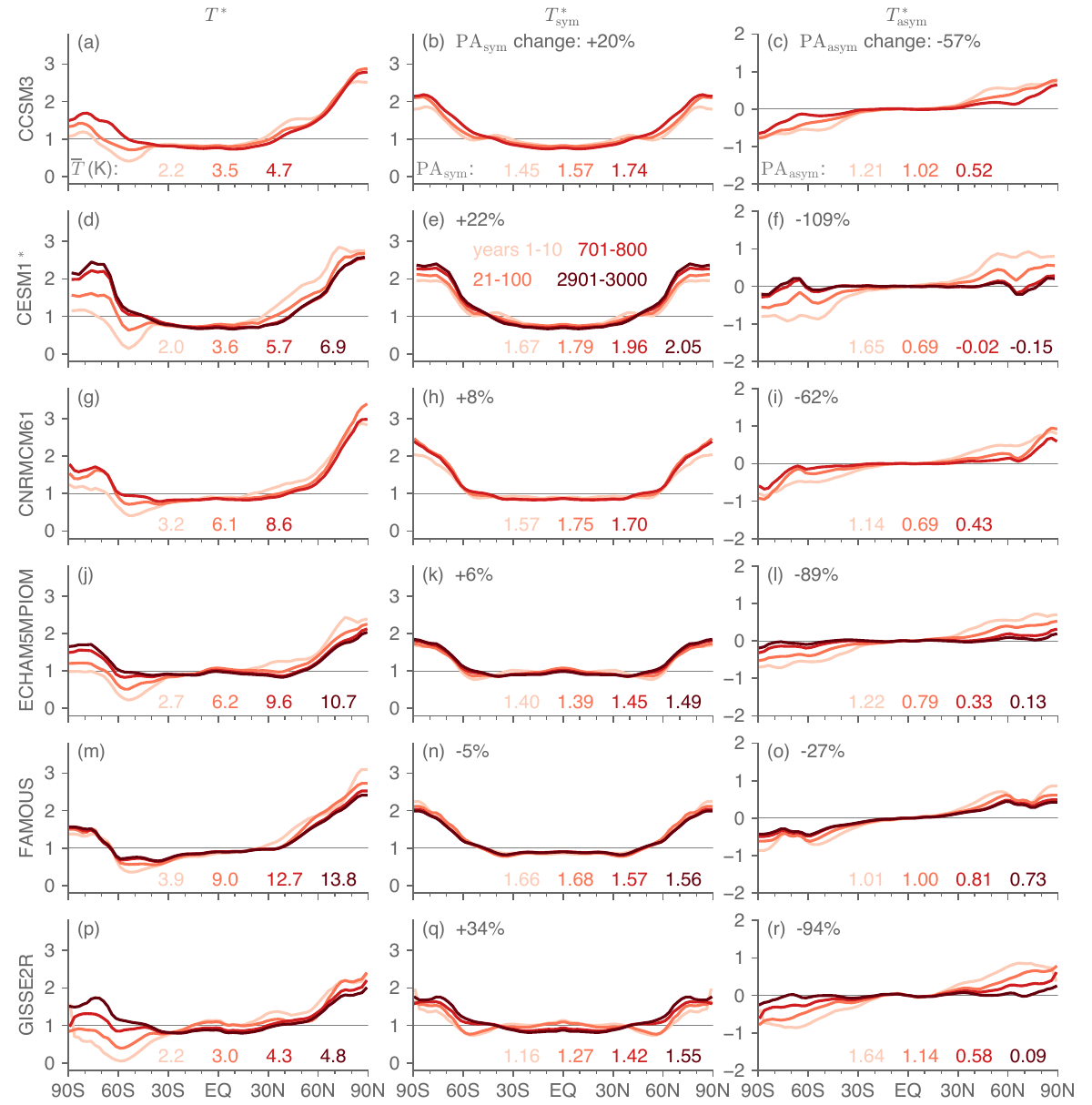}
\caption{For \ftcoo/ simulations in six of the twelve GCMs analyzed, (left) full, (middle) hemispherically symmetric component, and (right) antisymmetric component of zonal-mean surface air temperature change in years 1-10, 21-100, 701-800, and if available years 2901-3000, with each period as indicated in the legend in panel e.  Printed values at the bottom of each panel are: (left column), the mean warming during that period, (middle column) the symmetric polar amplification index for that period, and (right column) the northern minus southern hemisphere polar amplification index for that period.  Values at the top of the middle-column panels are the fractional change in the symmetric polar amplification index from the first to the last period, and in the right-column panels are the same but for the northern-southern difference.  For CESM1 (second row), the asterisk signifies that these simulations are not from LongRunMIP (unlike the eleven other models).   The remaining six models are shown in Fig.~\ref{fig:4x-last6}.}
\label{fig:4x-first6}
\end{figure*}

\begin{figure*}[h]
\centering
\includegraphics[width=\textwidth]{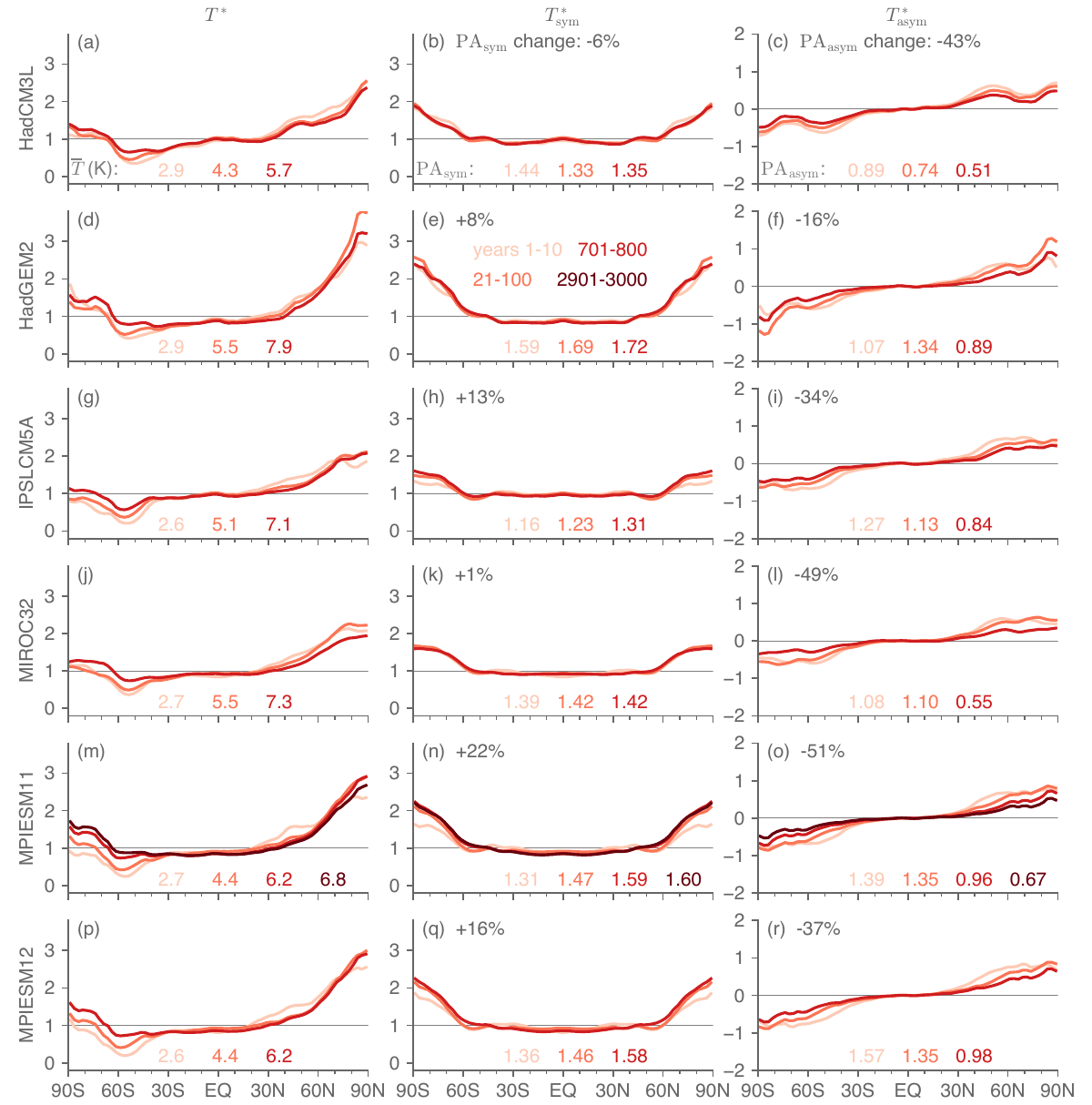}
\caption{Same as Fig.~\ref{fig:4x-first6}, but for the remaining six GCMs.}
\label{fig:4x-last6}
\end{figure*}

Fig.~\ref{fig:4x-first6} shows the \ftcoo/-forced \(T^*\), \(\Tsn\), and \(\Tan\) fields for each period specified above in six of the twelve GCMs; the remaining six are shown in Fig.~\ref{fig:4x-last6}.  Printed in each \(T^*\) panel are that model's global-mean warming (henceforth \(\olT\)) for each time period, in each \(\Tsn\) panel that model's \(\pasym\) for each period and its percentage change from the first decade to the last available period, and in each \(\Tan\) panel that model's \(\paasym\) for each period and its first-to-last percentage change.  Across models, \(\olT\) spans 2.0-3.9~K in the first decade and increases monotonically afterward in all models, with values 3.0-9.0~K in years 21-100, 4.3-12.7~K in years 701-800, and 4.8-13.8~K in years 2901-3000.  In the first decade only, in all twelve \(T^*\) minimizes over the Southern Ocean.  For nearly all models and latitudes south of \(\sim\)40\degr{}S, \(T^*\) increases monotonically in time.  The evolution of northern extratropical warming varies more across models, but in most \(T^*\) decreases with time over much of \(\sim\)40-70\degr{}N.

\(\Tsn\) is polar-amplified in all models and time periods, and by eye it either changes modestly or becomes somewhat more polar-amplified with time.
Quantitatively, \(\pasym\) spans 1.16-1.67 across models in the initial decade, 1.23-1.79 in years 21-100, 1.31-1.96 in years 701-800, and 1.49-2.05 in years 2901-3000.  In one outlier, GISSE2R, it increases from the first decade to the last available period by 34\%, and unique to this model the \(\Tsn\) field is similar for the first two periods but then seems to jump to one shared by the latter two periods.  In the remaining eleven GCMs, \(\pasym\) changes by \({\leq8}\%\) in six models (with negative change in FAMOUS, -5\%, and HadCM3L, -6\%), and increases by \({\leq}22\%\) in the remaining five.  As such, we consider a weak change to moderate increase in \(\pasym\) from decadal to millennial timescales under \ftcoo/ to be an empirically robust response across these models.

The \(\Tan\) field reflects greater Arctic than Antarctic warming initially but also a weakening in time of that difference in all models.  Quantitatively, \(\paasym\) is positive in the first decade in all models, spanning 0.89-1.65, and then decreases monotonically in ten of twelve models, spanning 0.69-1.35 in years 21-100, -0.02 to +0.98 in years 701-800, and -0.15 to +0.73 in years 2901-3000.  The two negative values, both in CESM1, indicate that Antarctic warming exceeds Arctic warming.  The fractional change in \(\paasym\) from the first decade to the last available period is negative in all models but varies considerably, -16\% to -109\%.  We consider a moderate to complete reduction in \(\paasym\) to likewise be a robust response.

Given these two robust responses, empirically each model's \(\pasym\) in the initial decade provides a nontrivial albeit approximate lower bound on its near-equilibrium \(\pasym\) value---and for models in which \(\Tan\) weakens strongly leaving \(T^*\approx\Tsn\) at near-equilibrium, this extends to polar amplification in each hemispheric cap.  By eye, indeed the full \(\Tsn\) field changes less with time in nearly all models than does \(T^*\) and less still than \(\Tan\).  We may quantify this using the ratio \(\paasym/\pasym\), which ranges from -0.01 in CESM1 to +0.64 in IPSLCM5A for years 701-800 and -0.07 in CESM1 to 0.47 in FAMOUS for years 2901-3000.  It becomes less positive between these periods in all five models run to years 2901-3000, at which point it is \(\leq10\%\) in magnitude in CESM1, ECHAM5MPIOM, and GISSE2R.  From Fig.~\ref{fig:4x-first6} the correspondence between the initial \(\Tsn\) and final \(T^*\) fields is debatable for the two whose \(\paasym/\pasym\) ratios are not small (0.42 in MPIESM11 and 0.47 in FAMOUS), intermediate for GISSE2R (recall its aforementioned jump in \(\Tsn\) after the first century), but clear for CESM1 and ECHAM5MPIOM.

Having established these behaviors in full-physics GCMs, we turn to better understanding them via two simpler models: first with an MEBM to clarify the physical mechanisms underlying the robust \(\Tsn\) and \(\Tan\) behaviors, and then with a box model to explore the GCM diversity in \(\pasym\) and \(\paasym\) evolutions.

\section{Moist Energy Balance Model}
\label{sec:mebm}

\begin{figure}[h]
\centering
\includegraphics[width=0.45\textwidth]{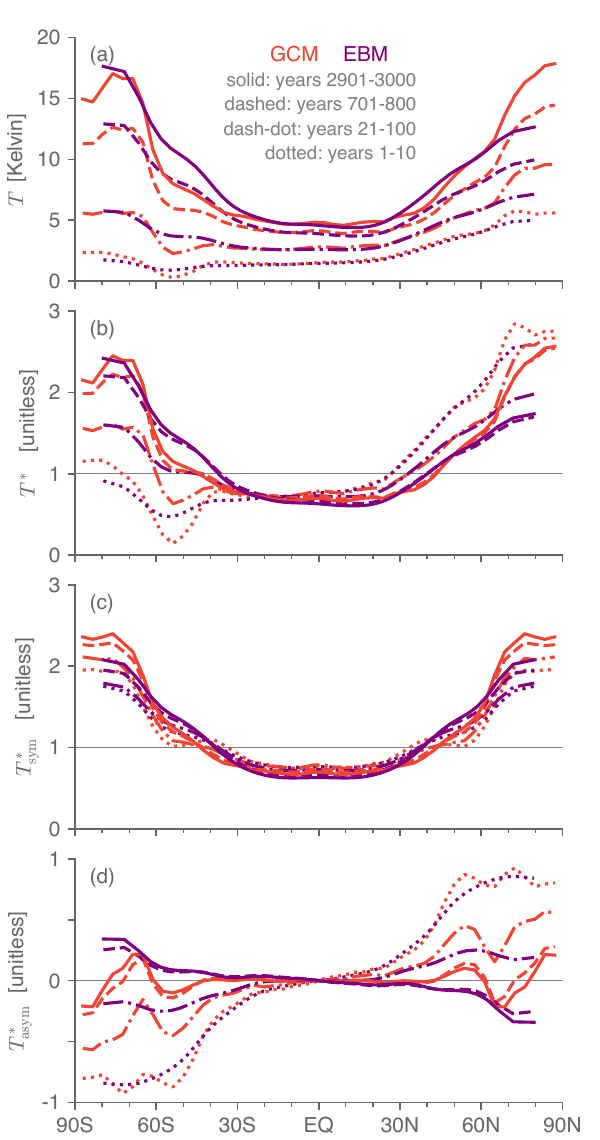}
\caption{Surface air temperature response in the CESM1 \ftcoo/ simulation at the four selected time periods and in the moist energy balance model simulations meant to reproduce the CESM1 \ftcoo/ simulation at each of those time periods, as indicated by the text in panel (a).  Panels from top to bottom show the temperature (a) raw (in units Kelvin), (b) mean-normalized (unitless), (c) mean-normalized symmetric component (unitless), and (d) mean-normalized antisymmetric component (unitless).  Note differing vertical axis spans in each panel.}
\label{fig:mebm}
\end{figure}

Fig.~\ref{fig:mebm} shows \(T\), \(T^*\), \(\Tsn\), and \(\Tan\) for each time period in the CESM1 \ftcoo/ simulation and the corresponding MEBM simulations; recall the MEBM simulations differ from one another only in the time period in CESM1 from which the radiative feedback parameter (\(\lambda\)) and ocean heat uptake (OHU; \(\ohu\)) were diagnosed.  The MEBM captures the mean warming (\(\olT\) is within 0.3~K of CESM1 for all four periods) and raw warming patterns reasonably well and therefore \(T^*\)---in particular the gradual, modest increase in polar amplification in \(\Tsn\) and the steady but severe weakening of \(\Tan\).  High-latitude warming gradients are insufficiently sharp \citep[a common feature of MEBMs with uniform diffusivity, \eg/][]{bonan_sources_2018}, yielding a slight low bias in the Arctic amplification index by years 2901-3000 (1.62 in the MEBM \vs/ 1.97 in CESM1) but less so for the Antarctic index due to compensating within-region \(T\) biases (2.16 in the MEBM \vs/ 2.13 in CESM1 for years 2901-3000).  More importantly, the MEBM suitably captures the fractional change in both \(\pasym\) and \(\paasym\) from CESM1: the MEBM \(\paasym\) decreases by -132\% (from 0.83 to -0.27, \vs/ -109\% in CESM1), and its \(\pasym\) increases by 19\% (from 1.59 to 1.89, \vs/ 22\% in CESM1).  As such, we can use the MEBM to further probe the underlying physical mechanisms.

\begin{figure}[h]
\centering
\includegraphics[width=0.5\textwidth]{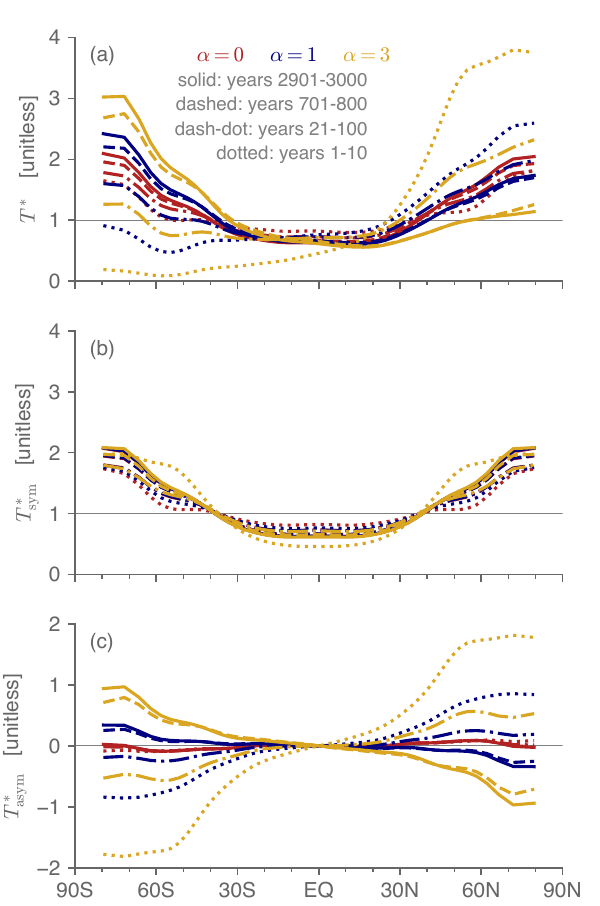}
\caption{Mean-normalized (a) full, (b) symmetric, and (c) antisymmetric surface air temperature anomaly fields in MEBM simulations with the antisymmetric components of the radiative feedback parameter and ocean heat uptake fields multiplied by the factor \(\alpha\), with red curves for \(\alpha=0\), blue for \(\alpha=1\) (\ie/ unchanged), and dark yellow for \(\alpha=3\).  Dotted, dash-dotted, dashed, and solid lines correspond to years 1-10, 21-100, 701-800, and 2901-3000 respectively of the CESM1.0.4 abrupt\ftcoo/ simulation.  Note that the vertical axis range is identical in panels (a) and (b) but not (c), while the vertical axis spacing is identical in all three panels.}
\label{fig:mebm-asymm}
\end{figure}

To test the role of antisymmetries in \(\lambda\) and \(\ohu\), Fig.~\ref{fig:mebm-asymm} shows \(T^*\), \(\Tsn\), and \(\Tan\) from simulations with \(\lambda\) and \(\ohu\) replaced by \(\lambda\sym\) and \(\ohu\sym\).  The resulting symmetric warming patterns (red curves) closely resemble the original ones (blue curves).\footnote{Modest antisymmetries stem from the radiative forcing and the climatological surface air temperature used to compute \(\partial_T q_\mr{sat}\); in additional simulations with these also symmetrized (not shown), the warming pattern is very similar to the symmetric pattern shown.}  Quantitatively, \(\olT\) changes from the full MEBM simulation by \(\leq\)0.3~K and \(\pasym\) by 0.04 or 3\% in all periods.  In a complementary simulation, the antisymmetric components are amplified rather than suppressed: we set \({\lambda^*=\lambda\sym+\alpha\lambda\asym}\), where \(\lambda^*\) is the modified feedback parameter, \(\ohu\) is modified likewise, and \({\alpha=3}\) (yellow curves in Fig.~\ref{fig:mebm-asymm}).  Although this strongly modifies \(T\) and \(\Tan\) in all periods, after the first decade \(\olT\) changes by at most 0.7~K, qualitatively the \(\Tsn\) pattern changes weakly, and quantitatively \(\pasym\) changes by \(\leq\)0.03.  These results suggest that \(\Tsn\) and \(\pasym\) are largely insensitive to \(\lambda\asym\) and \(\ohu\asym\), leading us to focus henceforth on \(\lambda\sym\) and \(\ohu\sym\).

\begin{figure}[h]
\centering
\includegraphics[width=0.45\textwidth]{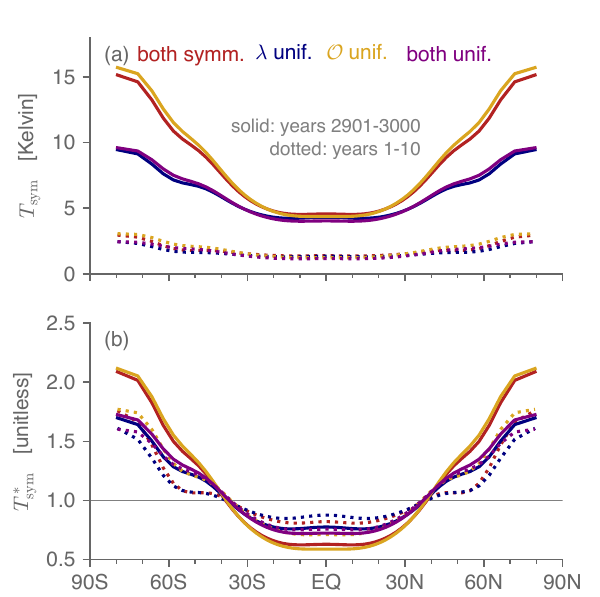}
\caption{Surface air temperature response in MEBM simulations corresponding to (dotted curves) years 1-10 and (solid curves) years 2901-3000 of the CESM1 \ftcoo/ simulation, shown either (a) raw (in units Kelvin) or (b) mean-normalized (unitless).  Colors correspond to simulations with (red) both \(\lambda\) and \(\ohu\) symmetrized but not uniform, (blue) \(\lambda\) set to its global-mean value at all latitudes, (yellow) \(\ohu\) set to its global-mean value at all latitudes, or (purple) both \(\ohu\) and \(\lambda\) set to their global-mean values at all latitudes.}
\label{fig:mebm-unif}
\end{figure}

To test the role of the spatial patterns of \(\lambda\sym\) and \(\ohu\sym\), Fig.~\ref{fig:mebm-unif} shows \(\Tsym\) and \(\Tsn\) in simulations with \(\lambda\) and \(\ohu\) replaced at all latitudes by their global averages, \(\ol\lambda\) and \(\ol\ohu\) respectively (purple curves).\footnote{In the MEBM, there is no true distinction between OHU and the radiative forcing, and so imposing the mean OHU uniformly can be equally conceptualized as reducing the global-mean radiative forcing.}  To focus on the biggest-picture behaviors, we restrict to simulations corresponding to the first decade (when \(\ol\lambda=-2.0\)~\wm/~K\inv{} and \(\ol\ohu=3.9\)~\wm/) and to years 2901-3000 (when \(\ol\lambda=-1.3\)~\wm/~K\inv{} and \(\ol\ohu=-0.4\)~\wm/).  With uniform \(\ohu\) and \(\lambda\), warming is reduced in the extratropics especially and globally averaged (\(\olT\) is 0.2~K cooler for the first decade and 1.7~K cooler for years 2901-3000), and \(T^*\) is less polar-amplified and changes less in time: \(\pasym\) is reduced from 1.54 to 1.51 (a 2\% decrease) in the first decade and from 1.89 to 1.60 (a 15\% decrease) in years 2901-3000.  In other words, the meridional patterns of \(\lambda\sym\) and \(\ohu\sym\) together make warming stronger in the mean and more polar-amplified compared to if they were uniform.

To distinguish the contributions of \(\lambda\sym\) \vs/ \(\ohu\sym\), Fig.~\ref{fig:mebm-unif} also includes simulations with only one of \(\lambda\) or \(\ohu\) made uniform.  For each time period, \(\Tsym\) of the original simulation more closely resembles the uniform-\(\ohu\) than uniform-\(\lambda\) simulation, which in turn more closely resembles the both-uniform simulation.  Quantitatively, in years 2901-3000 \(\olT\) is 0.2~K warmer than the \({\lambda=\lambda\sym}\), \({\ohu=\ohu\sym}\) case when \(\ohu\) is uniform \vs/ 1.7~K cooler when \(\lambda\) is uniform; \(\pasym\) is slightly increased with uniform \(\ohu\),
from 1.89 to 1.93 (2\%) \vs/ more strongly decreased with uniform \(\lambda\),
from 1.89 to 1.55 (-18\%).  In other words, the evolving spatial pattern of \(\lambda\sym\) (along with the mean of \(\ohu\)) acts to increase warming in the global mean and make it more polar-amplified, and these influences are stronger than those of the evolving spatial pattern of \(\ohu\sym\) (along with the mean of \(\lambda\)) in determining \(\Tsym\) and \(\Tsn\).

\begin{figure*}[h]
\centering
\includegraphics[width=\textwidth]{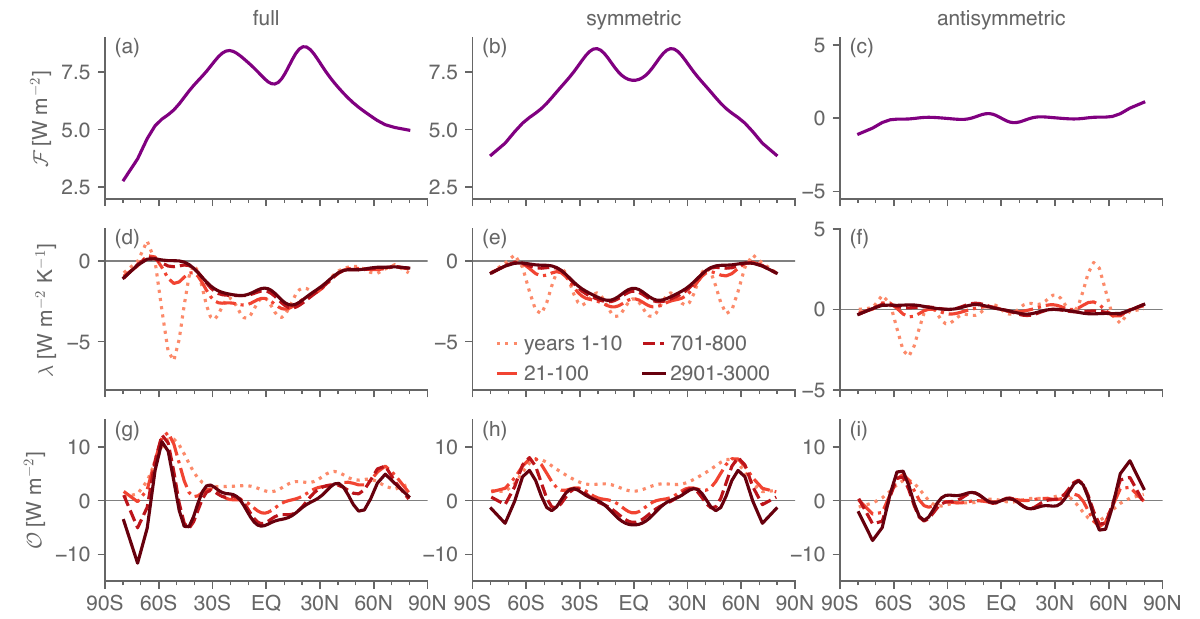}\\
\caption{From top to bottom, radiative forcing (\(\mathcal{F}\), units \wm/), radiative feedback parameter (\(\lambda\), units \wm/ K\inv{}), and ocean heat uptake (\(\ohu\), signed positive downward into the ocean, units \wm/) fields used for the MEBM simulations for each time period, with radiative forcing constant across time periods.  From left to right, panels show the full fields, the hemispherically symmetric component, and the hemispherically antisymmetric component.  For each row, the left and center panels have the same vertical axis ranges, and the right column has the same vertical axis spacing as the other columns but (except for the bottom row) not the same range.}
\label{fig:mebm-inputs}
\end{figure*}

To interpret this strong influence of \(\lambda\sym\), Fig.~\ref{fig:mebm-inputs} shows  \(\lambda\), \(\lambda\sym\), and \(\lambda\asym\) for each time period (as well as, for completeness in interpreting the various MEBM simulations, the corresponding \(\ohu\) fields and those of the time-invariant radiative forcing).  \(\lambda\) is negative at nearly all latitudes in all periods and is generally more negative in the tropics than high latitudes.  In the first decade it has a pronounced global minimum of \(\sim\)-6~\wm/~K\inv{} near 50\degr{}S, just equatorward of the global maximum in \(\ohu\) driven by Southern Ocean upwelling.
After the first decade, it becomes less stabilizing in the global average \citep[increasing climate sensitivity, \cf/][]{armour_time-varying_2013} and at most latitudes south of \(\sim\)10\degr{}N at least somewhat.  But the largest regional change is a vanishing of the sharp global minimum in \(\lambda\) by years 21-100.  With much weaker changes in the Northern Hemisphere, these signals project onto \(\lambda\sym\) as well.  \(\lambda\sym\) therefore becomes less stabilizing in the extratropics than tropics after the first decade, acting to increase polar amplification with time.\footnote{Strictly speaking, the \(\lambda\sym\) signal sits just outside our chosen Antarctic region boundary of 60\degr{}S. But the diffusive MSE transport in the MEBM clearly communicates this signal to the adjacent polar cap.}

Summarizing: the MEBM captures the CESM1 warming patterns reasonably well when forced with the latter's \(\lambda\) and \(\ohu\) fields; for \(\Tsym\) and \(\pasym\) to good approximation only the symmetric components of \(\lambda\) and \(\ohu\) matter; of these the predominant influence on the time evolution of \(\Tsym\) and \(\pasym\) is that of \(\lambda\sym\); and physically this stems from an initial strongly stabilizing radiative feedback over the Southern Ocean that vanishes after the first decade, making \(\lambda\sym\) become preferentially less stabilizing at high compared to low latitudes, increasing polar amplification.  We infer that changes in \(\Tsn\) are modest to the extent that changes in the spatial pattern of \(\lambda\sym\) are themselves modest.

\section{Box model of amplification indices}
\label{sec:box-model}

Having explored the mechanisms underlying the robust responses across the GCMs, we now investigate the cross-GCM discrepancies via a three-box, two-timescale model applied to two of the end-member GCMs noted above, CESM1 and FAMOUS \citep{smith_description_2008}.\footnote{C.f. Table~1 of \citet{rugenstein_equilibrium_2020}, FAMOUS is also an outlier in that the equilibrium warming estimated for \coo/-doubling differs nearly twofold whether \(2\times\) or \ftcoo/ simulations are used (4.40 and 8.55~K, respectively)}.  Our three-box model is an extension of the well-known two-timescale box model for global-mean warming \citep{held_probing_2010,geoffroy_transient_2013} to region-mean warming \citep[see also][]{geoffroy_pattern_2014} in the Arctic (60-90\degr{}N), Antarctic (60-90\degr{}S), and lower latitudes (60\degr{}S-60\degr{}N).  Recalling that CESM1 has the second-least mean warming, second-most positive change in \(\pasym\) (+22\%), and most negative change in \(\paasym\) (-109\%) whereas FAMOUS has the most mean warming, second-least positive (-5\%) change in \(\pasym\), and second-least negative (-27\%) change in \(\paasym\), this diagnosis of the regional warming timescales points toward potential causes of the spread in \(\Tsn\) and \(\Tan\) across the GCMs.

\begin{figure*}[h]
\centering
\includegraphics[width=\textwidth]{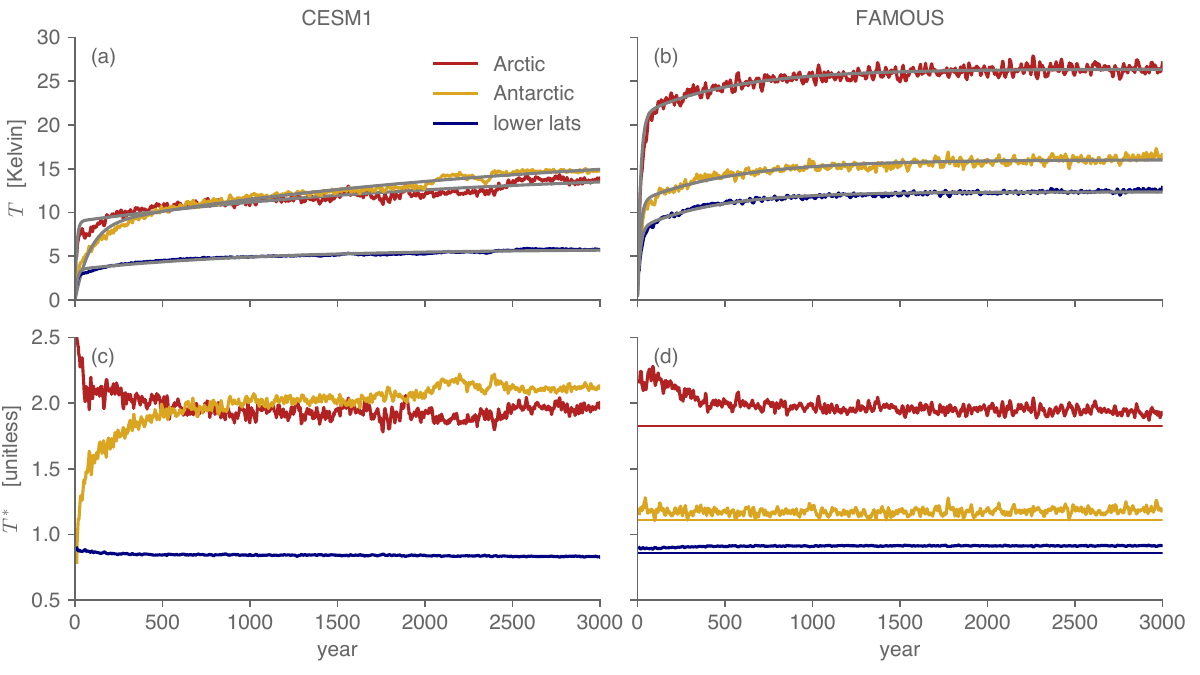}
\caption{Timeseries of 10-year running mean of (red) Arctic, (yellow) Antarctic, and (blue) low-latitude box-average temperatures in the abrupt \ftcoo/ simulation in (left) CESM1 and (right) FAMOUS.  Overlain gray curves are the fits from the simple two-layer box model for each region.  The top row shows raw fields, and the bottom row shows the same timeseries but each normalized by the global-mean warming.  Panel d also includes as thin horizontal lines the predictions from the box model under the approximation of horizontally uniform fast and slow warming timescales, as described in the text.}
\label{fig:box-model}
\end{figure*}

The two-timescale solution for a given region is given by
\begin{equation}
  \label{eq:2time}
T(t)=T_\mr{eq}\left[a_\mr{f}\left(1-e^{-t/\tau_\mr{f}}\right)+a_\mr{s}\left(1-e^{-t/\tau_\mr{s}}\right)\right],
\end{equation}
where \(T_\mr{eq}\) is the equilibrium temperature change, \(\tau_\mr{f}\) and \(\tau_\mr{s}\) are the fast and slow warming timescales respectively, and \(a_\mr{f}\) and \(a_\mr{s}\) are the fractional contributions of the fast and slow responses respectively to the equilibrium warming, with \(a_\mr{f}+a_\mr{s}=1\).  For each model and region, we fit \(T_\mr{eq}\), \(a_\mr{f}\), \(a_\mr{s}\), \(\tau_\mr{f}\), and \(\tau_\mr{s}\) of (\ref{eq:2time}) via nonlinear least-squares \citep[using the ``curve\_fit'' function of the scipy Python package;][]{virtanen_scipy_2020} applied to annual-mean timeseries of the region-mean surface temperature anomaly.  Although they do not appear explicitly, note that the ocean heat uptake and anomalous atmospheric energy flux divergence fields implicitly influence the values of all five parameters.  The resulting best-fit parameter values are listed in Table~\ref{table:box}, and the GCM regional-mean, annual-mean timeseries (smoothed via a 10-year running mean) and corresponding box-model solutions are shown both raw and mean-normalized in Fig.~\ref{fig:box-model}.

\begin{table*}
  \begin{center}
    \begin{tabular}{lrrrrr|rrrrr}
      \hline\hline
      &\multicolumn{5}{c}{CESM1}&\multicolumn{5}{c}{FAMOUS}\\
       &\(T_\mr{eq}\) [K] & \(\tau_\mr{f}\) [yr] & \(a_\mr{f}\) & \(\tau_\mr{s}\) [yr] & \(a_\mr{s}\) &\(T_\mr{eq}\) [K] & \(\tau_\mr{f}\) [yr] & \(a_\mr{f}\) & \(\tau_\mr{s}\) [yr] & \(a_\mr{s}\)\\
      Arctic     & 15.1 &  9.4 & 0.59 & 2223 & 0.41 & 26.4 & 15.9 & 0.78 & 471 & 0.22\\
      Antarctic  & 17.8 & 85.3 & 0.48 & 2564 & 0.52 & 16.1 & 15.3 & 0.71 & 588 & 0.29\\
      Lower lats &  5.8 & 12.7 & 0.60 & 1065 & 0.40 & 12.3 & 14.1 & 0.66 & 433 & 0.34\\
      Globe      &  7.1 & 15.9 & 0.58 & 1188 & 0.42 & 13.5 & 14.5 & 0.68 & 445 & 0.32\\
      \hline
    \end{tabular}
  \end{center}
  \caption{Best-fit values of the five parameters in the two-timescale model for the abrupt \ftcoo/ simulations in CESM1 and in FAMOUS for each of the three regions of our box model and for the global mean.  Units are Kelvin for \(T_\mr{eq}\) and years for \(\tau_\mr{f}\) and \(\tau_\mr{s}\); \(a_\mr{f}\) and \(a_\mr{s}\) are dimensionless.}
  \label{table:box}
\end{table*}

For CESM1, \(T_\mr{eq}\) is slightly higher for the Antarctic (17.8~K) than Arctic (15.1~K), both of which are \(\sim\)3\(\times\) higher than for lower latitudes (5.8~K).  The fast response timescales for the Arctic (9.4~yr) and lower latitudes (12.7~yr) are comparable and an order of magnitude less than the Antarctic timescale (85.3~yr).  Equilibrium warming is weighted fairly evenly between the fast and slow responses (\(a_\mr{f}=0.59\), 0.48, and 0.60 for the Arctic, Antarctic, and lower latitudes respectively).  The slow response timescales are all millennial---2,223, 2,564, and 1,065~yr respectively for the Arctic, Antarctic, and lower latitudes.  The two-timescale fit captures the overall evolution for each region fairly well, though with too sharp a shoulder after the initial decades for the Arctic and lower latitudes (Fig.~\ref{fig:box-model}a).  CESM1 also exhibits considerable centennial-timescale variability particularly after \(\sim\)1,800 yr \citep[roughly coinciding with the emergence of the Pacific Meridional Overturning Circulation in the Pliocene-like simulation;][]{burls_active_2017}.  For the fast response, the separation of the Antarctic timescale from the Arctic and lower latitudes is evident.  For the slow response, it is evident that both caps would continue warming nontrivially beyond year 3,000, which after all is only \(\sim\)1.2-1.3\(\times\) their slow response timescales.  The global-mean-normalized timeseries (Fig.~\ref{fig:box-model}c) show the initial strong Arctic amplification and the Antarctic subsequently catching up by around \(\sim\)500-600~yr.

The two-timescale fit is even better for FAMOUS than for CESM1 (Fig.~\ref{fig:box-model}b) and highlights the striking result that the Antarctic fast response timescale is slightly \emph{shorter} than the Arctic's---\(\tau_\mr{f}=14.1\), 15.3, and 15.9~yr respectively for lower-latitudes, Arctic, and Antarctic---unlike CESM1 and counter to physical intuition given the retarding influence of Southern Ocean upwelling.  The predicted equilibrium warming is over 10~K higher in the Arctic (26.4~K) than Antarctic (16.1~K), which in turn is less than 4~K warmer than the lower latitudes (12.3~K).  It is also weighted more toward the fast than slow response for all three regions, with \({a_\mr{f}=0.78}\), 0.71, and 0.66 respectively for the Arctic, Antarctic, and lower latitudes.  With comparable timescales and weightings for the fast response but much larger equilibrium warming in the Arctic, initial decades feature much greater Arctic than Antarctic warming.  The slow response timescale is similar for lower latitudes and Arctic (433 and 471 years respectively), and moderately longer for the Antarctic (588 years).  As such, the Antarctic continues warming somewhat longer than the rest of the globe, which moderately weakens the antisymmetric amplification.  Still, the Antarctic slow timescale is within  \(\sim\)25-35\% of the others.

These values motivate a first approximation for FAMOUS in which both timescales and their relative weights are uniform across regions.  Let \(T_\mr{N}\), \(T_\mr{S}\), and \(T_\mr{L}\) respectively be the Arctic, Antarctic, and low-latitude box temperature anomaly, and let \(\gamma\) be the ratio of the low-latitude surface area to the surface areas of either polar cap (with 60\degr{}S/N borders, \(\gamma\approx6.5\)).  Then \(\ol{T}=(T_\mr{S}+\gamma T_\mr{L}+T_\mr{S})/(\gamma+2)\), and the amplification indices are \(\pasym=(T_\mr{N}+T_\mr{S})/2\olT\) and \(\paasym=(T_\mr{N}-T_{\mr{S}})/2\olT\).  Denoting the equilibrium temperature anomalies \(T_{\mr{eq},i}\) for \(i\in\{\mr{N,L,S}\}\) and assuming each of \(\tau_\mr{f}\), \(\tau_\mr{s}\), \(a_\mr{f}\) and \(a_\mr{s}\) do not vary across the three boxes, the global-mean temperature anomaly is
\begin{align}
  \label{eq:box-famous-mean}
  \ol{T}(t)&=\dfrac{T_\mr{eq,S}+\gamma T_\mr{eq,L}+T_\mr{eq,N}}{\gamma+2}\,\times\\\nonumber
  &\left[a_\mr{f}(1-e^{-t/\tau_\mr{f}})+a_\mr{s}(1-e^{-t/\tau_\mr{s}})\right].
\end{align}
The global-mean-normalized temperature anomaly in each region is then
\begin{equation}
  \label{eq:box-famous-anom}
  \dfrac{T_i(t)}{\ol{T}(t)}=\dfrac{(\gamma+2)T_\mr{eq,i}}{T_\mr{eq,S}+\gamma T_\mr{eq,L}+T_\mr{eq,N}},\qquad i\in\{\mr{N,L,S}\}.
\end{equation}

This is independent of time.  Therefore so too are \(\pasym\) and \(\paasym\)---imperfect for the -27\% decrease in \(\paasym\) but capturing the modest -6\% change in \(\pasym\) well.  Fig.~\ref{fig:box-model}(c) shows the global-mean-normalized warming for each region for FAMOUS along with their predicted values from (\ref{eq:box-famous-anom}).  The simple approximation (\ref{eq:box-famous-anom}) is biased low for each region, but in reasonable agreement with (\ref{eq:box-famous-anom}) the FAMOUS timeseries vary modestly in time, at most for the Arctic by \(\sim\)10\% over the 3,000 years.

Summarizing, for CESM1 there are three relevant timescales.  In the initial decades, the Arctic warms rapidly but not the Antarctic, yielding large values of both \(\pasym\) and \(\paasym\).  The fast Antarctic warming transpires over subsequent decades to centuries, increasing \(\pasym\) but weakening \(\paasym\).  Over subsequent millennia, the slow responses emerge continuing to warm both polar caps, comparably to one another but more than lower latitudes, further increasing \(\pasym\) while decreasing \(\paasym\).  For FAMOUS, a surprisingly short timescale of Antarctic warming combined with much greater Arctic than Antarctic (or low-latitude) equilibrium warming combine to keep changes in both \(\pasym\) and \(\paasym\) moderate from decadal to millennial timescales.  These results show that the preferential initial Arctic \vs/ Antarctic amplification, though robust, can arise via rather different processes in different GCMs.

\section{Results across \coo/ levels and a Pliocene-like simulation in CESM1}
\label{sec:co2}

Though bounding a GCM's near-equilibrium \(\pasym\) from a short integration would be useful, our ultimate concern is what can be inferred for the real climate system, for which an instantaneous quadrupling of \coo/ is not directly relevant to anthropogenic warming---in which the \coo/ increase is gradual and (one dearly hopes) remains well below a quadrupling---nor those paleoclimate states for which non-\coo/ forcings are of first-order importance.  We therefore now present CESM1 simulations at 2-16\(\times\)\coo/ and the Pliocene-like simulation; these address the sensitivity of the results to \coo/ amount and to a strongly meridionally patterned, non-\coo/ forcing but do not directly address the issue of gradual rather than abrupt forcings, which we return to in the concluding discussion section below.

\begin{figure*}[h]
\centering
\includegraphics[width=\textwidth]{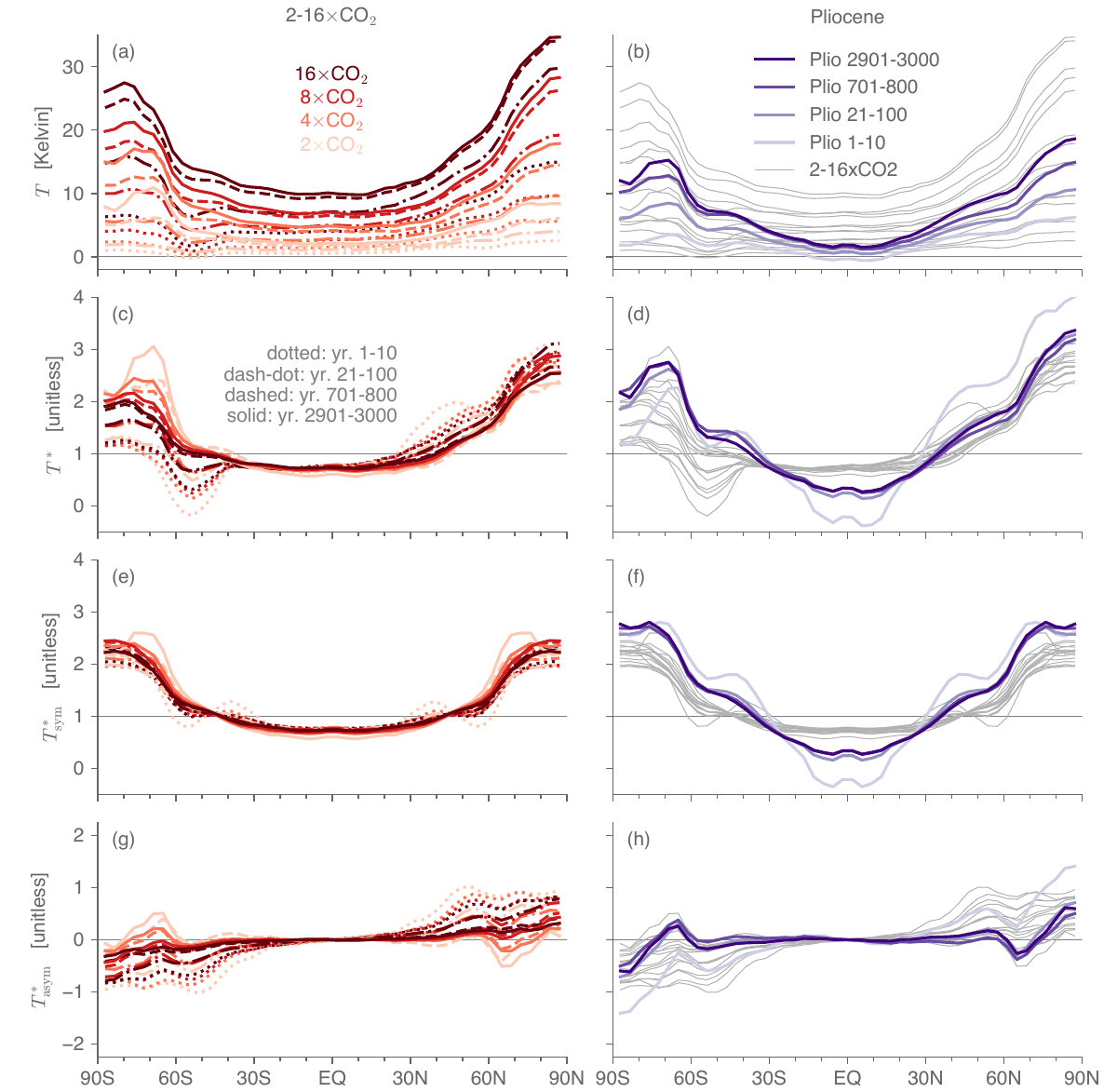}
\caption{Zonal-mean surface warming in (left column) 2, 4, 8, and 16\(\times\)\coo/ simulations and (right column) the Pliocene-like simulation in CESM1.  For \coo/, colors are according to the legend in panel a and line styles to the legend in panel c.  For the Pliocene-like simulation, the legend is in panel b.  Shown are (top row) raw warming in Kelvin, and the mean-normalized (next three rows) full field, symmetric component, and antisymmetric component.  Vertical axis range is identical in 2nd and 3rd row, which also have the same vertical axis spacing as the last row.}
  \label{fig:co2-plio}
\end{figure*}

The left column of Fig.~\ref{fig:co2-plio} shows \(T\), \(T^*\), \(\Tsn\), and \(\Tan\) for each perturbed \coo/ simulation and time period.  For \(T\), warming occurs at all latitudes, is weakest and relatively flat at low latitudes, and increases nearly monotonically moving from low to high latitudes (peaking in the Southern Hemisphere from \(\sim\)65\degr{}S for 2\(\times\)\coo/ to \(\sim\)80\degr{}S for 16\(\times\)\coo/).  Across \coo/ levels and time periods, low-latitude warming ranges from \(\sim\)2 to \(\sim\)11~K, peak SH high-latitude warming from \(\sim\)6 to \(\sim\)25~K, peak NH warming at the North Pole from \(\sim\)7 to \(\sim\)34~K.  \(\olT\) spans across periods 0.8-3.6~K, 2.0-6.9~K, 3.5-9.8~K, and 5.2-13.6~K respectively for 2, 4, 8, and \stcoo/.  For \(T^*\), the patterns are most similar across \coo/ amounts and timescales in the Tropics, moderately so in the northern extratropics, and least of all in the southern extratropics.  The Arctic amplification index decreases in time, and the Antarctic index increases, in all cases, both with the largest changes under \ttcoo/ (from 2.50 in the first decade to 1.97 for years 2901-3000 for the Arctic and from 0.75 to 2.57 for the Antarctic).

\(\Tsn\) is quite similar across \coo/ values and time periods, though least for the \ttcoo/ first and last periods.  For all \coo/ values the pattern becomes slightly more polar amplified in time, with low-latitude values decreasing and high-latitude values increasing.  Quantitatively, \(\pasym\) spans 1.62-2.29 across all \coo/ amounts and timescales (respectively occurring in years 1-10 and 2901-3000 under \ttcoo/).  In other words, \(\pasym\) starts smallest and ends up largest in \ttcoo/, increasing less with time (from the first to last period, by +41, 22, 18, and 8\% for 2-16\(\times\) respectively), and to a smaller near-equilibrium value (2.29, 2.05, 1.96, and 1.83 respectively) as \coo/ increases.  \(\Tan\) varies appreciably across the four timescales, reflecting the gradual catching-up of Antarctic and Southern Ocean warming with (and for 2 and \ftcoo/, surpassing) the initially rapid Arctic warming.  Similar to \(\pasym\) but with signs reversed, \(\paasym\) is most positive in the first decade under \ttcoo/ but then becomes most negative at near-equilibrium (0.9 and -0.3 in the first decade and years 2901-3000 respectively), and it decreases less with time as \coo/ increases (from the first to last period, by -132, -109, -81, and -72\% for 2-16\(\times\) respectively).  Particularly in the first century, \(\Tan\) groups together more by timescale than by \coo/ value (compare curves with the same line markings to those with the same color).  This likely reflects the intrinsic timescales of the underlying physical processes---no matter how large a radiative forcing, prevailing Southern Ocean upwelling inhibits initial local surface warming, while the deep ocean equilibration that acts to homogenize subsurface warming between the hemispheres takes millennia.



The right column of Fig.~\ref{fig:co2-plio} shows \(T\), \(T^*\), \(\Tsn\), and \(\Tan\) in the Pliocene-like simulation, with the corresponding 2-16\(\times\)\coo/ values underlain for comparison.  Recalling that cloud albedo is increased 15\degr{}S-15\degr{}N and decreased poleward thereof, the warming is unsurprisingly more polar-amplified in all time periods than under \coo/.  Mean warming is 1.6, 3.2, 4.6, and 5.5~K respectively in the four periods.  The first decade's mean-normalized fields, with \(\Tsn\) particularly cool at low latitudes and \(\Tan\) large near the poles, sit separate from the three subsequent periods across which they are very similar.  As for \coo/, the Arctic amplification index is initially large (3.2) and the Antarctic index smaller (1.9), but by years 21-100 they are nearly the same: 2.4, 2.2, and 2.4 for the Arctic in the three time periods \vs/ 2.3, 2.4, and 2.4 respectively.  \(\Tsn\) is more polar-amplified than the \coo/ cases, but like the \coo/ cases it changes modestly in time.  Quantitatively, \(\pasym=2.6\) in the first decade and 2.4 in the remaining three periods, and \(\paasym=0.6\) in the first decade and essentially vanishes (within -0.1 to +0.1) thereafter.

Summarizing, across \coo/ values in CESM1 the \(\Tsn\) pattern is qualitatively consistent, more so than \(\Tan\) due to relative Arctic warming increasing with \coo/.  Quantitatively, \(\pasym\) moderately increase with time for each \coo/ amount, but less so as \coo/ increases.  Similarly, \(\paasym\) decreases (that is, becomes more negative) in time at all \coo/ amounts, but less so as \coo/ increases.  Under the polar-amplified forcing of the Pliocene-like simulation, unsurprisingly surface warming is itself more polar-amplified than for the quasi-uniform \coo/ forcing, but it changes weakly after the first decade.  These results help to contextualize the cross-model spread in \(\pasym\) and \(\paasym\) under \ftcoo/.  They suggest that the fractional changes in \(\pasym\) and \(\paasym\) depend to a nontrivial extent on the forcing magnitude itself---consider that CESM1's 41\% increase in \(\pasym\) under \ttcoo/ is appreciably larger than all twelve GCMs under \ftcoo/ (at most +34\%).  Similarly, the 72-132\% range in the decrease of \(\paasym\) for CESM1 across \coo/ amounts is not vastly smaller than the range across LongRunMIP models under \ftcoo/ of 16-109\%.  At the same time, consistency of \(\pasym\) and \(\paasym\) at least after the first decade is even stronger under more meridionally structured forcing.




\section{Conclusions}
\label{sec:conc}

\subsection{Summary}

We decompose the zonal-mean surface air temperature response to abrupt \coo/ quadrupling from decadal to millennial timescales into hemispherically symmetric and antisymmetric components in twelve GCMs---eleven from LongRunMIP \citep{rugenstein_longrunmip_2019} plus a low-resolution version of CESM1.0.4.  Normalized by the contemporaneous global-mean warming, the symmetric warming component at a given time differs considerably across GCMs but for a given GCM changes modestly with time; a symmetric polar amplification index changes from the first decade to years 701-800 or (if available) years 2901-3000 by -6\% to +8\% in six of twelve, increases by 34\% in one outlier, and increases by 13-22\% in the remaining five.  The antisymmetric component weakens in time in all twelve, but this varies considerably across GCMs---near-equilibrium warming is appreciably antisymmetric in some including FAMOUS \vs/ almost entirely symmetric in some including CESM1.  Based on these results, we consider a weak change to moderate increase in symmetric polar amplification and a moderate to complete reduction in antisymmetric polar amplification to be robust responses and subsequently attempt to better understand them.

An MEBM prescribed with ocean heat uptake and radiative feedback parameter (\(\lambda\)) fields inferred from four different time periods of the \ftcoo/ CESM1 simulation captures the salient GCM behaviors.  In additional MEBM simulations with the antisymmetric components of \(\lambda\) and ocean heat uptake either removed or amplified, despite the antisymmetric warming pattern changing drastically, the symmetric warming pattern hardly changes.  Conversely, removing the meridional structure in \(\lambda\) (even with its global mean unchanged) causes three key changes: it reduces mean warming at each time period, it makes the warming pattern more polar-amplified at each time period, and it weakens the increase in time of the symmetric polar amplification index.  Of these three, the first two are a straightforward consequence of \(\lambda\) being less stabilizing overall at high than low latitudes.  The latter, we argue, results from the loss following the initial decade in a deep global minimum in \(\lambda\) just equatorward of the Southern Ocean.  This imprints onto the symmetric component of \(\lambda\), and with no comparable change at lower latitudes, the result is that radiative restoring becomes comparatively less stabilizing in the extratropics than tropics, promoting polar amplification.

To clarify causes of differences across the GCMs, a simple three-box, two-timescale model of warming in the Arctic, Antarctic, and lower-latitude sectors was fitted to 3,000-yr timeseries of annual-mean surface warming in the end-member models CESM1 and FAMOUS.  Strikingly, in FAMOUS there is effectively no difference in the fast response timescale between the Antarctic and elsewhere.  This runs counter to CESM1 where the Antarctic timescale is an order of magnitude larger and to physical intuition given the delaying effect of Southern Ocean upwelling.  Fortuitously however, it enables an analytical approximate solution that yields a time-invariant symmetric polar amplification index in good agreement with the GCM.

Finally, we investigate the sensitivity of these behaviors to the radiative forcing via additional CESM1 simulations.  The normalized symmetric warming pattern varies moderately across \coo/ magnitudes from 2-16\(\times\) preindustrial, with symmetric polar amplification increasing---and antisymmetric polar amplification decreasing---less in time the higher \coo/ is.  At least after the first decade, both components change even less in time in a simulation generating an early Pliocene-like surface climate attained through meridionally patterned cloud albedo perturbations.  Thus, qualitatively the symmetric component is insensitive in time and to forcing magnitude for a given forcing structure despite, unsurprisingly, depending sensitively on the forcing structure.

\subsection{Discussion}

Does the hemispherically symmetric/antisymmetric decomposition of polar-amplified warming add value over more conventional analyses?  In terms of the bulk amplification indices defined as the ratio of polar cap-averaged warming to globally averaged warming, admittedly the results are mixed.  Across the twelve GCMs under abrupt \ftcoo/, the Arctic amplification index spans 1.79-2.50 in the first decade and changes afterward by -0.53 to +0.04---a larger range than that of \(\pasym\) (-0.10 to +0.40) in absolute terms but actually smaller in percentage terms (-23\% to +2\% for the Arctic \vs/ -6 to +34\% for \(\pasym\)).  The Antarctic amplification index in the first decade spans 0.33-1.15 and changes thereafter from +0.04 to +1.27 (fractionally, +4 to +352\%), a larger range than the first-to-last-period change in \(\paasym\) of -0.09 to -0.90 (fractionally, -17\% to -109\%).  As such, a complementary view would be that the robust responses are a weak change to modest decrease in Arctic amplification and a weak to large increase in Antarctic amplification.  In either case, it is clear that model diversity in evolution of warming patterns across timescales is greatest for the southern extratropics, weakest in the tropics, and intermediate for the northern extratropics.

Further support for the southern high latitudes figuring centrally in model disagreement comes from the three-box, two-timescale model fitted to the two end members FAMOUS and CESM1.  Their most salient discrepancies are the \(\sim\)5.5-fold longer Antarctic fast-response timescale in CESM1 and the \(\sim\)2.5-4.5\(\times\) longer slow-response timescales in all regions for CESM1.  Both involve ocean dynamical processes---prevailing Southern Ocean upwelling and deep ocean equilibration---implying a predominant role for ocean model formulation.

Nevertheless, going beyond scalar amplification indices to the full latitude-by-latitude pattern, by eye from Fig.~\ref{fig:4x-first6},~\ref{fig:4x-last6}, and \ref{fig:co2-plio} clearly the mean-normalized symmetric component persists more in time from its values in initial decades over subsequent centuries and millennia than either the full field or its antisymmetric counterpart---and likewise (in CESM1 at least) across \coo/ concentrations.  We therefore argue that the hemispherically symmetric/antisymmetric decomposition merits further study.

One useful next step would be extending the analyses to additional, higher resolution, and more modern GCMs via the CMIP6 abrupt \ftcoo/ simulations---consider that the end-members CESM1 and FAMOUS are both low-resolution and/or simplified versions of CMIP5-class GCMs.  The CMIP abrupt \ftcoo/ runs typically span 150 years, precluding direct investigation of the multi-centennial and longer timescale behaviors.  However, for the twelve GCMs we analyze, values in years 21-100 are well correlated with those for years 701-800 for global mean warming (\({r=0.99}\)), Arctic amplification (\({r=0.94}\)), Antarctic amplification (\({r=0.78}\)), \(\pasym\) (\({r=0.87}\)), and \(\paasym\) (\({r=0.84}\)).

Though constraining global-mean warming \(\olT\) at any given timescale is not our main focus, we find noteworthy its spread across the twelve GCMs.  Near-equilibrium warming in the least sensitive model (GISSE2R) is surpassed within decades in half of the models.  Conversely, the most sensitive model (FAMOUS) warms more in the first century than ten of the other eleven do by years 701-800 and three of the other four do by years 2901-3000.  In addition, the warming in the initial decade is well correlated with the millennial-timescale warming: \({r=0.88}\) between \(\olT\) in years 1-10 \vs/ 701-800, and \({r=0.91}\) for years 1-10 \vs/ 2901-3000---raising the prospect of constraining future mean warming by its rapidity on decadal timescales.

The MEBM simulations suggest that, in CESM1 at least, \(\Tsn\) is largely time invariant after the first decade because the spatial stucture of the radiative feedback parameter itself hardly varies after the first decade.  How relevant this is to the real climate depends on the ability of GCMs to represent the presence (or in the case of CESM1, absence) of any state-dependent feedbacks that could result in regional changes in the radiative feedback parameter in time as mean warming increases.  At least in the MEBM, making the radiative feedback parameter uniform---even with its global mean intact---considerably reduces climate sensitivity.  It is well known that the evolving feedback field tends to increase sensitivity with time, but this is often understood in regards to its global-mean value becoming less stabilizing.  In other words, the MEBM suggests that state-dependent climate sensitivity is related to having non uniform feedbacks.

The ultimate motivation for this work is to infer as much as possible regarding anthropogenic climate change in the real climate system from limited data records.  What can be inferred from the real climate system based on these results, bracketing temporarily questions of validity?  Historical radiative forcing is characterized by two factors we have yet to consider.  First is a gradual rather than abrupt \coo/ increase.  A useful starting place would be standard 1\% per year \coo/ increase simulations.  In initial years to decades when the radiative forcing is still relatively small, global-mean warming will likely be too small for the mean-normalized fields to be meaningful, and the spatial pattern of warming will likely be strongly influenced by internal variability.  For that reason, analyzing these fields in one or more available large ensembles could be useful.  Second is a complex spatiotemporal evolution with nontrivial antisymmetric component owing to anthropogenic aerosols, volcanoes, and land-use change.  We have not examined forcings with large hemispherically antisymmetric components, and it is possible that the \(\Tsn\) and \(\Tan\) behaviors under such forcings would differ from the robust behaviors we have shown under predominantly symmetric forcing.

The 2-\stcoo/ simulation results from CESM1 suggest that, in that model at least, the \(\Tsn\) and \(\pasym\) behaviors are reasonably insensitive to \coo/ values ranging from 2-16\(\times\) preindustrial, though with \(\pasym\) moderately increasing with \coo/, while \(\Tan\) and \(\paasym\) become more weighted to the Arctic as \coo/ increases.  This helps contextualize the near vanishing of \(\Tan\) and \(\pasym\) under \ftcoo/: this is not intrinsic to \coo/-forced warming; rather \ftcoo/ happens to be the amount at which the processes controlling the difference between the caps have comparable strengths.  The \ttcoo/ simulation is the only one in which southern hemisphere sea ice does not disappear entirely; under \stcoo/ it is nearly gone by years 21-100, under \etcoo/ by years 701-800, and under \ftcoo/ by years 2901-3000 (not shown).  Given the importance of sea ice loss in the severity of polar amplification \citep[\eg/][]{dai_arctic_2019}, this likely contributes to the \ttcoo/ being least like the others regarding the fields of our interest.  \citet{heede_time_2020}, analyzing nearly identical abrupt \coo/-increase simulations (in the same model, though run for only 500~years and with five ensemble members each), discuss in detail the ways in which the \ttcoo/ simulation differs from the larger-magnitude \coo/ increase simulations.

We conclude by noting that this analysis depended crucially on the existence of an ensemble of millennial-scale integrations in LongRunMIP \citep{rugenstein_longrunmip_2019}, and that, in attempting to bound longer-timescale behaviors from shorter integrations, it complements ongoing efforts such as the ``fast-forward'' technique \citep{saintmartin_fast_2019} to attain equilibrium solutions in GCMs more rapidly.

\appendix[A]
\appendixtitle{Moist energy balance model formulation}

For the \coo/ radiative forcing (\(\mathcal{F}(\lat)\)) we use the spatially varying instantaneous forcing of \citet{huang_inhomogeneous_2016} computed for a doubling of \coo/.  This will not be identical to the radiative forcing computed with our particular GCM due to dependencies on the climatology \citep[\eg/][]{merlis_direct_2015,huang_pattern_2017}.
We convert this instantaneous \ttcoo/ radiative forcing into a stratosphere-adjusted \ftcoo/ radiative forcing by doubling it and then adding 2.4~\wm/ uniformly to yield a conventional global-mean value of 7.0~\wm/.  It is shown in Fig.~\ref{fig:mebm-inputs}(a) and is identical across all MEBM simulations presented.

The feedback parameter \(\lambda\) is diagnosed  this radiative forcing field and fields taken from the GCM \ftcoo/ simulation:
\begin{equation}
  \label{eq:lambda}
  \lambda(\lat) = -\dfrac{\mathcal{F}(\lat) - \mathcal{T}(\lat)}{T_\mr{gcm}(\lat)},
\end{equation}
where \(\mathcal{T}\) is the anomalous TOA radiative flux in the GCM (signed positive downward), and \(T_\mr{gcm}\) is the anomalous surface air temperature in the GCM.  The ``gcm'' subscript is meant to emphasize that the temperature field in the denominator of (\ref{eq:lambda}) is that diagnosed from the GCM, not the MEBM's own computed temperature (whereas the temperature field that \(\lambda\) multiplies in (\ref{eq:mebm}) is that of the MEBM).  One MEBM simulation is performed for each of the four time periods of interest, each with \(\mathcal{T}\), \(T_\mr{gcm}\), and \(\ohu\) taken from the CESM1 \ftcoo/ simulation averaged over that time period.

For the diffusive approximation to atmospheric energy transport convergence, because all quantities are anomalies, surface MSE is linearized as \(h = T (1 + \mathcal{H} L \partial_T q_\mr{sat} )/c_p\), with relative humidity \(\mathcal{H}\), saturation specific humidity \(q_\mr{sat}\), and latent heat of vaporization \(L\).  The partial derivative of the saturation vapor pressure, \(\partial_T q_\mr{sat}\), is evaluated using the zonal-mean climatological surface air temperature from the GCM averaged over years 701-800 of the control simulation.  The parameter values for all constant coefficients are standard: \(\mathcal{H} = 0.8\), \(\mathcal{D} = 0.3 \, \mathrm{W \, m^{-2} \, K^{-1}}\), \(c_p = 1004.6 \, \mathrm{J \, kg^{-1} \, K^{-1}}\), and \(L = 2.5 \times 10^6 \, \mathrm{J \, kg^{-1}}\).

The MEBM is integrated to equilibrium using a fourth-order Runge-Kutta timestepping scheme.  A second-order finite difference scheme is used for the \(\nabla^2\) operator \citep{wagner_how_2015}.  There are 60 model grid points evenly spaced in \(\sinlat=1/30\) increments, with gridpoint centers in each hemisphere from \(\sinlat\approx0.12\) (corresponding to \(\lat\approx4.8^\circ\)) to \(\sinlat\approx0.98\) (corresponding to \(\lat\approx79.5^\circ\)).  The CESM1 fields, which are evenly spaced in latitude over 48 boxes spanning \(\sim\)1.8- 87.2\degr{} in each hemisphere with \(\sim\)3.6\degr{} spacing, are spectrally transformed at order 20 to the MEBM grid.

\acknowledgments
We thank William Wang for generating several figures that facilitated our analyses.  S.A.H. was supported during different periods of this study by an NSF Atmospheric and Geospace Sciences Postdoctoral Research Fellowship (NSF Award \#1624740), a Caltech Foster and Coco Stanback Postdoctoral Fellowship, and a Columbia University Earth Institute Fellowship.  T.M.M acknowledges support from NSERC.  N.J.B. acknowledges support from NSF Award \#1844380 and is supported by the Alfred P. Sloan Foundation as a Research Fellow.  A.V.F. acknowledges support from the ARCHANGE project (ANR-18-MPGA-0001, France).  Three anonymous reviewers provided extremely useful comments and motivated the analyses of LongRunMIP and the box model.

\bibliographystyle{ametsoc2014}
\bibliography{./references}

\begin{thebibliography}{48}
\providecommand{\natexlab}[1]{#1}
\providecommand{\url}[1]{\texttt{#1}}
\renewcommand{\UrlFont}{\rmfamily}
\providecommand{\urlprefix}{URL }
\expandafter\ifx\csname urlstyle\endcsname\relax
  \providecommand{\doi}[1]{doi:\discretionary{}{}{}#1}\else
  \providecommand{\doi}{doi:\discretionary{}{}{}\begingroup
  \urlstyle{rm}\Url}\fi
\providecommand{\eprint}[2][]{\url{#2}}

\bibitem[{Alexeev and Jackson(2012)Alexeev, and Jackson}]{alexeev_polar_2012}
Alexeev, V.~A., and C.~H. Jackson, 2012: Polar amplification: Is atmospheric
  heat transport important? \textit{Clim Dyn}, \textbf{41~(2)}, 533--547,
  \doi{10.1007/s00382-012-1601-z}.

\bibitem[{Andrews et~al.(2015)Andrews, Gregory,, and
  Webb}]{andrews_dependence_2015}
Andrews, T., J.~M. Gregory, and M.~J. Webb, 2015: The {{Dependence}} of
  {{Radiative Forcing}} and {{Feedback}} on {{Evolving Patterns}} of {{Surface
  Temperature Change}} in {{Climate Models}}. \textit{J. Climate},
  \textbf{28~(4)}, 1630--1648, \doi{10.1175/JCLI-D-14-00545.1}.

\bibitem[{Armour et~al.(2013)Armour, Bitz,, and Roe}]{armour_time-varying_2013}
Armour, K.~C., C.~M. Bitz, and G.~H. Roe, 2013: Time-{{Varying Climate
  Sensitivity}} from {{Regional Feedbacks}}. \textit{J. Climate},
  \textbf{26~(13)}, 4518--4534, \doi{10.1175/JCLI-D-12-00544.1}.

\bibitem[{Armour et~al.(2019)Armour, Siler, Donohoe,, and
  Roe}]{armour_meridional_2019}
Armour, K.~C., N.~Siler, A.~Donohoe, and G.~H. Roe, 2019: Meridional
  {{Atmospheric Heat Transport Constrained}} by {{Energetics}} and {{Mediated}}
  by {{Large-Scale Diffusion}}. \textit{J. Climate}, \textbf{32~(12)},
  3655--3680, \doi{10.1175/JCLI-D-18-0563.1}.

\bibitem[{Bonan et~al.(2018)Bonan, Armour, Roe, Siler,, and
  Feldl}]{bonan_sources_2018}
Bonan, D.~B., K.~C. Armour, G.~H. Roe, N.~Siler, and N.~Feldl, 2018: Sources of
  {{Uncertainty}} in the {{Meridional Pattern}} of {{Climate Change}}.
  \textit{Geophys. Res. Lett.}, \textbf{45~(17)}, 9131--9140,
  \doi{10.1029/2018GL079429}.

\bibitem[{Burls and Fedorov(2014{\natexlab{a}})Burls, and
  Fedorov}]{burls_simulating_2014}
Burls, N.~J., and A.~V. Fedorov, 2014{\natexlab{a}}: Simulating {{Pliocene}}
  warmth and a permanent {{El Ni\~no-like}} state: {{The}} role of cloud
  albedo. \textit{Paleoceanography}, \textbf{29~(10)}, 893--910,
  \doi{10.1002/2014PA002644}.

\bibitem[{Burls and Fedorov(2014{\natexlab{b}})Burls, and
  Fedorov}]{burls_what_2014}
Burls, N.~J., and A.~V. Fedorov, 2014{\natexlab{b}}: What {{Controls}} the
  {{Mean East}}\textendash{{West Sea Surface Temperature Gradient}} in the
  {{Equatorial Pacific}}: {{The Role}} of {{Cloud Albedo}}. \textit{J.
  Climate}, \textbf{27~(7)}, 2757--2778, \doi{10.1175/JCLI-D-13-00255.1}.

\bibitem[{Burls et~al.(2017)Burls, Fedorov, Sigman, Jaccard, Tiedemann,, and
  Haug}]{burls_active_2017}
Burls, N.~J., A.~V. Fedorov, D.~M. Sigman, S.~L. Jaccard, R.~Tiedemann, and
  G.~H. Haug, 2017: Active {{Pacific}} meridional overturning circulation
  ({{PMOC}}) during the warm {{Pliocene}}. \textit{Science Advances},
  \textbf{3~(9)}, e1700\,156, \doi{10.1126/sciadv.1700156}.

\bibitem[{Dai et~al.(2019)Dai, Luo, Song,, and Liu}]{dai_arctic_2019}
Dai, A., D.~Luo, M.~Song, and J.~Liu, 2019: Arctic amplification is caused by
  sea-ice loss under increasing {{CO}} 2. \textit{Nat Commun}, \textbf{10~(1)},
  1--13, \doi{10.1038/s41467-018-07954-9}.

\bibitem[{Danabasoglu and Gent(2009)Danabasoglu, and
  Gent}]{danabasoglu_equilibrium_2009}
Danabasoglu, G., and P.~R. Gent, 2009: Equilibrium {{Climate Sensitivity}}:
  {{Is It Accurate}} to {{Use}} a {{Slab Ocean Model}}? \textit{J. Climate},
  \textbf{22~(9)}, 2494--2499, \doi{10.1175/2008JCLI2596.1}.

\bibitem[{Ding et~al.(2014)Ding, Wallace, Battisti, Steig, Gallant, Kim,, and
  Geng}]{ding_tropical_2014}
Ding, Q., J.~M. Wallace, D.~S. Battisti, E.~J. Steig, A.~J.~E. Gallant, H.-J.
  Kim, and L.~Geng, 2014: Tropical forcing of the recent rapid {{Arctic}}
  warming in northeastern {{Canada}} and {{Greenland}}. \textit{Nature},
  \textbf{509~(7499)}, 209--212, \doi{10.1038/nature13260}.

\bibitem[{Dong et~al.(2020)Dong, Armour, Zelinka, Proistosescu, Battisti,
  Zhou,, and Andrews}]{dong_intermodel_2020}
Dong, Y., K.~C. Armour, M.~D. Zelinka, C.~Proistosescu, D.~S. Battisti,
  C.~Zhou, and T.~Andrews, 2020: Intermodel {{Spread}} in the {{Pattern
  Effect}} and {{Its Contribution}} to {{Climate Sensitivity}} in {{CMIP5}} and
  {{CMIP6 Models}}. \textit{Journal of Climate}, \textbf{33~(18)}, 7755--7775,
  \doi{10.1175/JCLI-D-19-1011.1}.

\bibitem[{Fedorov et~al.(2015)Fedorov, Burls, Lawrence,, and
  Peterson}]{fedorov_tightly_2015}
Fedorov, A.~V., N.~J. Burls, K.~T. Lawrence, and L.~C. Peterson, 2015: Tightly
  linked zonal and meridional sea surface temperature gradients over the past
  five million years. \textit{Nature Geosci}, \textbf{8~(12)}, 975--980,
  \doi{10.1038/ngeo2577}.

\bibitem[{Feldl et~al.(2017)Feldl, Anderson,, and
  Bordoni}]{feldl_atmospheric_2017}
Feldl, N., B.~T. Anderson, and S.~Bordoni, 2017: Atmospheric {{Eddies Mediate
  Lapse Rate Feedback}} and {{Arctic Amplification}}. \textit{J. Climate},
  \textbf{30~(22)}, 9213--9224, \doi{10.1175/JCLI-D-16-0706.1}.

\bibitem[{Flannery(1984)}]{flannery_energy_1984}
Flannery, B.~P., 1984: Energy {{Balance Models Incorporating Transport}} of
  {{Thermal}} and {{Latent Energy}}. \textit{J. Atmos. Sci.}, \textbf{41~(3)},
  414--421, \doi{10.1175/1520-0469(1984)041<0414:EBMITO>2.0.CO;2}.

\bibitem[{Frierson and Hwang(2012)Frierson, and
  Hwang}]{frierson_extratropical_2012}
Frierson, D. M.~W., and Y.-T. Hwang, 2012: Extratropical {{Influence}} on
  {{ITCZ Shifts}} in {{Slab Ocean Simulations}} of {{Global Warming}}.
  \textit{J. Climate}, \textbf{25~(2)}, 720--733,
  \doi{10.1175/JCLI-D-11-00116.1}.

\bibitem[{Geoffroy and {Saint-Martin}(2014)Geoffroy, and
  {Saint-Martin}}]{geoffroy_pattern_2014}
Geoffroy, O., and D.~{Saint-Martin}, 2014: Pattern decomposition of the
  transient climate response. \textit{Tellus A: Dynamic Meteorology and
  Oceanography}, \textbf{66~(1)}, 23\,393, \doi{10.3402/tellusa.v66.23393}.

\bibitem[{Geoffroy et~al.(2013)Geoffroy, {Saint-Martin}, Olivi{\'e}, Voldoire,
  Bellon,, and Tyt{\'e}ca}]{geoffroy_transient_2013}
Geoffroy, O., D.~{Saint-Martin}, D.~J.~L. Olivi{\'e}, A.~Voldoire, G.~Bellon,
  and S.~Tyt{\'e}ca, 2013: Transient {{Climate Response}} in a {{Two-Layer
  Energy-Balance Model}}. {{Part I}}: {{Analytical Solution}} and {{Parameter
  Calibration Using CMIP5 AOGCM Experiments}}. \textit{Journal of Climate},
  \textbf{26~(6)}, 1841--1857, \doi{10.1175/JCLI-D-12-00195.1}.

\bibitem[{Heede et~al.(2020)Heede, Fedorov,, and Burls}]{heede_time_2020}
Heede, U.~K., A.~V. Fedorov, and N.~J. Burls, 2020: Time {{Scales}} and
  {{Mechanisms}} for the {{Tropical Pacific Response}} to {{Global Warming}}:
  {{A Tug}} of {{War}} between the {{Ocean Thermostat}} and {{Weaker Walker}}.
  \textit{Journal of Climate}, \textbf{33~(14)}, 6101--6118,
  \doi{10.1175/JCLI-D-19-0690.1}.

\bibitem[{Held et~al.(2010)Held, Winton, Takahashi, Delworth, Zeng,, and
  Vallis}]{held_probing_2010}
Held, I.~M., M.~Winton, K.~Takahashi, T.~Delworth, F.~Zeng, and G.~K. Vallis,
  2010: Probing the {{Fast}} and {{Slow Components}} of {{Global Warming}} by
  {{Returning Abruptly}} to {{Preindustrial Forcing}}. \textit{Journal of
  Climate}, \textbf{23~(9)}, 2418--2427, \doi{10.1175/2009JCLI3466.1}.

\bibitem[{Henry et~al.(2021)Henry, Merlis, Lutsko,, and
  Rose}]{henry_decomposing_2021}
Henry, M., T.~M. Merlis, N.~J. Lutsko, and B.~E.~J. Rose, 2021: Decomposing the
  {{Drivers}} of {{Polar Amplification}} with a {{Single-Column Model}}.
  \textit{Journal of Climate}, \textbf{34~(6)}, 2355--2365,
  \doi{10.1175/JCLI-D-20-0178.1}.

\bibitem[{Huang et~al.(2016)Huang, Tan,, and Xia}]{huang_inhomogeneous_2016}
Huang, Y., X.~Tan, and Y.~Xia, 2016: Inhomogeneous radiative forcing of
  homogeneous greenhouse gases. \textit{Journal of Geophysical Research:
  Atmospheres}, \textbf{121~(6)}, 2780--2789, \doi{10.1002/2015JD024569}.

\bibitem[{Huang et~al.(2017)Huang, Xia,, and Tan}]{huang_pattern_2017}
Huang, Y., Y.~Xia, and X.~Tan, 2017: On the pattern of {{CO2}} radiative
  forcing and poleward energy transport. \textit{J. Geophys. Res. Atmos.},
  \textbf{122~(20)}, 2017JD027\,221, \doi{10.1002/2017JD027221}.

\bibitem[{Hwang et~al.(2011)Hwang, Frierson,, and Kay}]{hwang_coupling_2011}
Hwang, Y.-T., D.~M.~W. Frierson, and J.~E. Kay, 2011: Coupling between
  {{Arctic}} feedbacks and changes in poleward energy transport.
  \textit{Geophysical Research Letters}, \textbf{38~(17)},
  \doi{10.1029/2011GL048546}.

\bibitem[{Jansen et~al.(2018)Jansen, Nadeau,, and
  Merlis}]{jansen_transient_2018}
Jansen, M.~F., L.-P. Nadeau, and T.~M. Merlis, 2018: Transient versus
  {{Equilibrium Response}} of the {{Ocean}}'s {{Overturning Circulation}} to
  {{Warming}}. \textit{J. Climate}, \textbf{31~(13)}, 5147--5163,
  \doi{10.1175/JCLI-D-17-0797.1}.

\bibitem[{Li et~al.(2013)Li, {von Storch},, and Marotzke}]{li_deep-ocean_2013}
Li, C., J.-S. {von Storch}, and J.~Marotzke, 2013: Deep-ocean heat uptake and
  equilibrium climate response. \textit{Clim Dyn}, \textbf{40~(5)}, 1071--1086,
  \doi{10.1007/s00382-012-1350-z}.

\bibitem[{Manabe et~al.(1991)Manabe, Stouffer, Spelman,, and
  Bryan}]{manabe_transient_1991}
Manabe, S., R.~J. Stouffer, M.~J. Spelman, and K.~Bryan, 1991: Transient
  {{Responses}} of a {{Coupled Ocean}}\textendash{{Atmosphere Model}} to
  {{Gradual Changes}} of {{Atmospheric CO2}}. {{Part I}}. {{Annual Mean
  Response}}. \textit{Journal of Climate}, \textbf{4~(8)}, 785--818,
  \doi{10.1175/1520-0442(1991)004<0785:TROACO>2.0.CO;2}.

\bibitem[{Marshall et~al.(2015)Marshall, Scott, Armour, Campin, Kelley,, and
  Romanou}]{marshall_oceans_2015}
Marshall, J., J.~R. Scott, K.~C. Armour, J.-M. Campin, M.~Kelley, and
  A.~Romanou, 2015: The ocean's role in the transient response of climate to
  abrupt greenhouse gas forcing. \textit{Clim Dyn}, \textbf{44~(7-8)},
  2287--2299, \doi{10.1007/s00382-014-2308-0}.

\bibitem[{Merlis(2015)}]{merlis_direct_2015}
Merlis, T.~M., 2015: Direct weakening of tropical circulations from masked
  {{CO2}} radiative forcing. \textit{PNAS}, \textbf{112~(43)},
  13\,167--13\,171, \doi{10.1073/pnas.1508268112}.

\bibitem[{Merlis and Henry(2018)Merlis, and Henry}]{merlis_simple_2018}
Merlis, T.~M., and M.~Henry, 2018: Simple {{Estimates}} of {{Polar
  Amplification}} in {{Moist Diffusive Energy Balance Models}}. \textit{J.
  Climate}, \textbf{31~(15)}, 5811--5824, \doi{10.1175/JCLI-D-17-0578.1}.

\bibitem[{Previdi et~al.(2020)Previdi, Janoski, Chiodo, Smith,, and
  Polvani}]{previdi_arctic_2020}
Previdi, M., T.~P. Janoski, G.~Chiodo, K.~L. Smith, and L.~M. Polvani, 2020:
  Arctic {{Amplification}}: {{A Rapid Response}} to {{Radiative Forcing}}.
  \textit{Geophysical Research Letters}, \textbf{47~(17)}, e2020GL089\,933,
  \doi{10.1029/2020GL089933}.

\bibitem[{Roe et~al.(2015)Roe, Feldl, Armour, Hwang,, and
  Frierson}]{roe_remote_2015}
Roe, G.~H., N.~Feldl, K.~C. Armour, Y.-T. Hwang, and D.~M.~W. Frierson, 2015:
  The remote impacts of climate feedbacks on regional climate predictability.
  \textit{Nature Geosci}, \textbf{8~(2)}, 135--139, \doi{10.1038/ngeo2346}.

\bibitem[{Rose et~al.(2014)Rose, Armour, Battisti, Feldl,, and
  Koll}]{rose_dependence_2014}
Rose, B. E.~J., K.~C. Armour, D.~S. Battisti, N.~Feldl, and D.~D.~B. Koll,
  2014: The dependence of transient climate sensitivity and radiative feedbacks
  on the spatial pattern of ocean heat uptake. \textit{Geophysical Research
  Letters}, \textbf{41~(3)}, 1071--1078, \doi{10.1002/2013GL058955}.

\bibitem[{Rugenstein et~al.(2019)}]{rugenstein_longrunmip_2019}
Rugenstein, M., and Coauthors, 2019: {{LongRunMIP}}: {{Motivation}} and
  {{Design}} for a {{Large Collection}} of {{Millennial-Length AOGCM
  Simulations}}. \textit{Bull. Amer. Meteor. Soc.}, \textbf{100~(12)},
  2551--2570, \doi{10.1175/BAMS-D-19-0068.1}.

\bibitem[{Rugenstein et~al.(2020)}]{rugenstein_equilibrium_2020}
Rugenstein, M., and Coauthors, 2020: Equilibrium {{Climate Sensitivity
  Estimated}} by {{Equilibrating Climate Models}}. \textit{Geophysical Research
  Letters}, \textbf{47~(4)}, e2019GL083\,898, \doi{10.1029/2019GL083898}.

\bibitem[{Russotto and Biasutti(2020)Russotto, and
  Biasutti}]{russotto_polar_2020}
Russotto, R.~D., and M.~Biasutti, 2020: Polar amplification as an inherent
  response of a circulating atmosphere: Results from the {{TRACMIP}}
  aquaplanets. \textit{Geophysical Research Letters}, \textbf{n/a~(n/a)},
  e2019GL086\,771, \doi{10.1029/2019GL086771}.

\bibitem[{Saint-Martin et~al.(2019)}]{saintmartin_fast_2019}
Saint-Martin, D., and Coauthors, 2019: Fast forward to perturbed equilibrium
  climate. \textit{Geophysical Research Letters}, \textbf{0~(ja)},
  \doi{10.1029/2019GL083031}.

\bibitem[{Senior and Mitchell(2000)Senior, and
  Mitchell}]{senior_time-dependence_2000}
Senior, C.~A., and J.~F.~B. Mitchell, 2000: The time-dependence of climate
  sensitivity. \textit{Geophysical Research Letters}, \textbf{27~(17)},
  2685--2688, \doi{10.1029/2000GL011373}.

\bibitem[{Shields et~al.(2012)Shields, Bailey, Danabasoglu, Jochum, Kiehl,
  Levis,, and Park}]{shields_low-resolution_2012}
Shields, C.~A., D.~A. Bailey, G.~Danabasoglu, M.~Jochum, J.~T. Kiehl, S.~Levis,
  and S.~Park, 2012: The {{Low-Resolution CCSM4}}. \textit{J. Climate},
  \textbf{25~(12)}, 3993--4014, \doi{10.1175/JCLI-D-11-00260.1}.

\bibitem[{Shin and Kang(2021)Shin, and Kang}]{shin_how_2021}
Shin, Y., and S.~M. Kang, 2021: How {{Does}} the {{High-Latitude Thermal
  Forcing}} in {{One Hemisphere Affect}} the {{Other Hemisphere}}?
  \textit{Geophysical Research Letters}, \textbf{48~(24)}, e2021GL095\,870,
  \doi{10.1029/2021GL095870}.

\bibitem[{Smith et~al.(2008)Smith, Gregory,, and
  Osprey}]{smith_description_2008}
Smith, R.~S., J.~M. Gregory, and A.~Osprey, 2008: A description of the
  {{FAMOUS}} (version {{XDBUA}}) climate model and control run.
  \textit{Geoscientific Model Development}, \textbf{1~(1)}, 53--68,
  \doi{10.5194/gmd-1-53-2008}.

\bibitem[{Stephens et~al.(2015)Stephens, O'Brien, Webster, Pilewski, Kato,, and
  Li}]{stephens_albedo_2015}
Stephens, G.~L., D.~O'Brien, P.~J. Webster, P.~Pilewski, S.~Kato, and J.-l. Li,
  2015: The albedo of {{Earth}}. \textit{Rev. Geophys.}, \textbf{53~(1)},
  2014RG000\,449, \doi{10.1002/2014RG000449}.

\bibitem[{Stuecker et~al.(2018)}]{stuecker_polar_2018}
Stuecker, M.~F., and Coauthors, 2018: Polar amplification dominated by local
  forcing and feedbacks. \textit{Nature Climate Change}, \textbf{8~(12)}, 1076,
  \doi{10.1038/s41558-018-0339-y}.

\bibitem[{Tebaldi and Arblaster(2014)Tebaldi, and
  Arblaster}]{tebaldi_pattern_2014}
Tebaldi, C., and J.~M. Arblaster, 2014: Pattern scaling: {{Its}} strengths and
  limitations, and an update on the latest model simulations. \textit{Climatic
  Change}, \textbf{122~(3)}, 459--471, \doi{10.1007/s10584-013-1032-9}.

\bibitem[{Virtanen et~al.(2020)}]{virtanen_scipy_2020}
Virtanen, P., and Coauthors, 2020: {{SciPy}} 1.0: Fundamental algorithms for
  scientific computing in {{Python}}. \textit{Nat Methods}, \textbf{17~(3)},
  261--272, \doi{10.1038/s41592-019-0686-2}.

\bibitem[{Voigt et~al.(2013)Voigt, Stevens, Bader,, and
  Mauritsen}]{voigt_observed_2013}
Voigt, A., B.~Stevens, J.~Bader, and T.~Mauritsen, 2013: The {{Observed
  Hemispheric Symmetry}} in {{Reflected Shortwave Irradiance}}. \textit{J.
  Climate}, \textbf{26~(2)}, 468--477, \doi{10.1175/JCLI-D-12-00132.1}.

\bibitem[{Voigt et~al.(2014)Voigt, Stevens, Bader,, and
  Mauritsen}]{voigt_compensation_2014}
Voigt, A., B.~Stevens, J.~Bader, and T.~Mauritsen, 2014: Compensation of
  {{Hemispheric Albedo Asymmetries}} by {{Shifts}} of the {{ITCZ}} and
  {{Tropical Clouds}}. \textit{Journal of Climate}, \textbf{27~(3)},
  1029--1045, \doi{10.1175/JCLI-D-13-00205.1}.

\bibitem[{Wagner and Eisenman(2015)Wagner, and Eisenman}]{wagner_how_2015}
Wagner, T. J.~W., and I.~Eisenman, 2015: How {{Climate Model Complexity
  Influences Sea Ice Stability}}. \textit{J. Climate}, \textbf{28~(10)},
  3998--4014, \doi{10.1175/JCLI-D-14-00654.1}.

\end{thebibliography}

\end{document}